\documentclass[12pt,journal,draftclsnofoot,onecolumn]{IEEEtran}
\normalsize

\ifCLASSINFOpdf
\else
\fi

\usepackage{mathtools}

\usepackage{cite}
\usepackage{amsmath,amssymb,amsfonts}
\usepackage{textcomp}
\usepackage{eqnarray, bm}
\usepackage[pdftex]{graphicx}
\usepackage{algorithmic}
\usepackage{algorithm}
\usepackage{multicol}
\usepackage{graphicx}
\usepackage{lipsum, color}
\usepackage{mathtools, cuted, nccmath}
\usepackage{commath}

\usepackage{epstopdf}
\usepackage{subfigure}
\usepackage{mathtools, nccmath}
\usepackage{epsfig}
\DeclareGraphicsExtensions{.eps}
\usepackage{stfloats}
\usepackage{physics}

\usepackage{subfig}
\allowdisplaybreaks[4]

\DeclareMathOperator*{\argmax}{argmax}

\usepackage{xcolor, soul}
\sethlcolor{red}

\usepackage [english]{babel}
\usepackage [autostyle, english = american]{csquotes}
\MakeOuterQuote{"}

\begin{document}
\title{Computation Offloading for IoT in C-RAN: Optimization and Deep Learning}

\author{Chandan~Pradhan,~\IEEEmembership{Student Member,~IEEE,}
       Ang~Li,~\IEEEmembership{Member,~IEEE,}
       Changyang~She,~\IEEEmembership{Member,~IEEE,}
       Yonghui~Li,~\IEEEmembership{Fellow,~IEEE,}
       and~Branka~Vucetic,~\IEEEmembership{Fellow,~IEEE \vspace{-5.10ex}}% <-this % stops a space
\thanks{Chandan Pradhan, Ang Li, Changyang She, Yonghui Li and Branka Vucetic are  with  the  Centre  of  Excellence  in  Telecommunications, School of Electrical and Information Engineering, University of Sydney, Sydney, NSW 2006, Australia (e-mail: \{chandan.pradhan, ang.li2, changyang.she, yonghui.li, branka.vucetic\}@sydney.edu.au).}}

\maketitle

\begin{abstract}

We consider computation offloading for Internet-of-things (IoT) applications in  multiple-input-multiple-output (MIMO) cloud-radio-access-network (C-RAN). Due to the limited battery life and computational capability in the IoT devices (IoTDs), the computational tasks of the IoTDs are offloaded  to a MIMO C-RAN, where a MIMO radio resource head (RRH) is connected to a baseband unit (BBU) through a capacity-limited fronthaul link, facilitated by the spatial filtering and uniform scalar quantization. We formulate a computation offloading optimization problem to minimize the total transmit power of the IoTDs while satisfying the latency requirement of the computational tasks, and find that the problem is non-convex. To obtain a feasible solution, firstly the spatial filtering matrix is locally optimized at the MIMO RRH. Subsequently, we leverage the alternating optimization framework for joint optimization on the residual variables at the BBU, where the baseband combiner is obtained in a closed-form, the resource allocation sub-problem is solved through successive inner convexification, and the number of quantization bits is obtained by a line-search method. As a low-complexity approach, we deploy a supervised deep learning method, which is trained with the solutions to our optimization algorithm. Numerical results validate the effectiveness of the proposed algorithm and the deep learning method.

\end{abstract}
%\vspace{-2mm}


\IEEEpeerreviewmaketitle

\section{Introduction}

Internet-of-things (IoT) has a great potential to impact our lives in the future by providing solutions related to multiple sectors of industry, smart homes, transportation, etc. It is predicted that there will be about 50 billion IoT devices by 2020 \cite{popli2019survey, feltrin2019narrowband}. The deployment of  a large-scale IoT ecosystem requires the IoT devices (IoTDs) with a small physical size to be built from cost-efficient hardware components, which results in major challenges due to  their limited battery life and computational capability.  More importantly, IoT applications require flexibility in handling diverse latency requirements \cite{popli2019survey}. To address the limited battery life and computational capability, the computational task in an IoTD can be migrated to a more powerful server \cite{Kumar2013}, which is known as \textit{computation offloading} \cite{Kumar2013, she2019cross}. Furthermore, technologies like massive  multiple-input-multiple-output (MIMO) \cite{ngo2013energy, luMIMOOverview2014, LISWaad2018} and cloud-radio-access-network (C-RAN) \cite{zhang2015CRAN,  Barbarossa2013, Barbarossa2015, jointKim2019, liu2019TwoScale, li2018minmax, jointBckAcce2015} can be exploited to augment the process of computation offloading and manage the corresponding latency requirement imposed by the IoT applications. 

Massive MIMO, characterized by the deployment of a huge number of antennas, is a key enabling technique for 5G wireless systems \cite{ngo2013energy, luMIMOOverview2014}.  More recently, \textit{extra-large scale MIMO (xL-MIMO)} as a step further has received increasing research attention \cite{LISWaad2018, de2019non, amiriXtra2018, BeyondMIMO2018}. In XL-MIMO, a large antenna array in the order of hundreds and thousands is integrated into a large man-made structure, for example, walls of buildings in the residential rooms, airports, or large shopping malls, as a scaled-up version of the massive MIMO systems where the spatial dimension provides an additional degree of freedom to further enhance the performance of the massive MIMO systems \cite{Mar2014, de2019non}. In addition, xL-MIMO systems provide a better coverage with a line-of-sight (LOS) channel, which also simplifies the corresponding channel estimation \cite{LISWaad2018, amiriXtra2018, BeyondMIMO2018}. However, the implementation of such xL-MIMO systems is challenging due to the deployment complexity along with the increasing requirement for the baseband signal processing, which is proportional to the number of antenna elements.

C-RAN can be a potential technique to  overcome the above challenges for the xL-MIMO systems. Specifically, C-RAN migrates the baseband signal processing to a baseband unit (BBU) that is equipped with a powerful server in the "cloud", while the radio frequency (RF) functionalities are implemented at the remote radio head (RRH) \cite{CRanSurvey2014}. By combining massive MIMO with C-RAN, the deployment complexity of the conventional massive MIMO systems can be greatly alleviated, since only analog components such as  antennas and RF chains with a limited signal processing capability are required \cite{zhang2015CRAN,   liu2019TwoScale}. However, moving the signal processing of a massive MIMO system from the RRH to the central BBU requires a huge amount of digitally sampled  data to be transmitted over the fronthaul link. Therefore, it is necessary to compress the uplink data at the RRH to satisfy the capacity limit of the fronthaul link. Accordingly, in \cite{zhang2015CRAN},  the authors proposed a data compression method which reduces the dimension of the signals received across the multiple antennas through spatial filtering, followed by a uniform scalar quantization across the reduced dimension. To further reduce the cost and power consumption of the hardware components in a C-RAN system, hybrid analog-digital designs have subsequently been applied to the massive MIMO  C-RANs  \cite{jointKim2019, yu2016alternating, sohrabi2016hybrid, liu2019TwoScale}, where the number of RF chains at the RRH can be reduced.

Different from the wireless systems in \cite{zhang2015CRAN} and \cite{liu2019TwoScale} where the uplink communication has a high spectral efficiency requirement, the latency-constrained IoT applications pursue low data rates with a higher energy efficiency performance while meeting their stringent latency constraints. In this regard, computation offloading with the massive MIMO C-RAN can be leveraged by allocating  the transmit power and computational resource at the BBU server to each IoTD while satisfying their latency requirement. While there have been studies on admission control and offloading strategies for computation offloading in a single-antenna wireless network in \cite{addOffload2018, offStrategy2019} and references therein, there are only a limited number of works in the literature that study the joint communication and computational resource  allocation for computation offloading in a MIMO C-RAN \cite{Barbarossa2013, Barbarossa2015, li2018minmax}. In \cite{Barbarossa2013} and \cite{Barbarossa2015}, the offloading problems were formulated to minimize the total transmit power and energy consumption of the devices, respectively, while meeting the latency constraints. The computation offloading method proposed in \cite{li2018minmax} aimed to minimize the maximum latency of all the devices.  Nevertheless, these works did not consider the compression and quantization of the received signal at the RRH as in \cite{zhang2015CRAN, liu2019TwoScale}, which can lead to infeasible data traffic for a capacity-limited fronthaul link. Moreover, the latency incurred at the capacity-limited fronthaul link in transferring the data from the RRH to the BBU can critically impede the execution of the computational tasks. Accordingly, the above mentioned drawbacks call for the joint design of the resource allocations, the compression and quantization strategies, especially for the latency-critical IoT applications.

Furthermore, the iterative nature of the  solutions proposed in \cite{Barbarossa2013, Barbarossa2015, li2018minmax, zhang2015CRAN, liu2019TwoScale} are computationally demanding for real-time implementation, especially when the computational tasks have a stringent latency requirement. Recently, deep learning has become a promising tool in solving difficult wireless communication problems, such as resource allocation \cite{dong2019deep, pcDeep2018}, channel decoding \cite{decode2018} and channel estimation \cite{DeepUSVXu2019, DeepSVKim2018}, which can return a near-optimal solution with a low-complexity implementation. The main idea of deep learning is to treat a given computationally expensive algorithm as a "black box", and try to learn the policy obtained with the algorithm by using a \textit{deep neural network} (DNN)  \cite{SunDeep2018, luong2019applications}. Specifically, the authors in \cite{SunDeep2018} have shown that the DNN can be trained to learn the non-linear mapping between the input and output of an algorithm, where the outputs obtained from running the algorithm offline can be used as labeled samples to train the DNN. Accordingly, the trained DNN only requires simple operations such as  matrix-vector multiplications to obtain near-optimal solutions.

Motivated by the above, in this work we consider the computation offloading problem for the IoTDs in a massive MIMO C-RAN deployed in an indoor environment, where multiple receive antennas that are spread across one of the walls act as an xL-MIMO RRH. Specifically, the uplink signals, encoding the computation bits from the IoTDs, are firstly received at the xL-MIMO before spatial filtering. Subsequently, the filtered signals are quantized \cite{LiuJointPower2015} and transmitted to the BBU via a capacity-limited fronthaul link, where a baseband combiner corresponding to each IoTD extracts and forwards the respective signal to the BBU server. We focus on the minimization of the total transmit power of the IoTDs, while satisfying the latency requirement of their corresponding computational task. We summarize the main contributions of the paper below:

\begin{enumerate}

\item We establish a computation offloading optimization problem to minimize the total transmit power of the IoTDs by jointly optimizing the communication and computational resource allocation policy, the spatial filtering design at the xL-MIMO RRH, the number of quantization bits, and the baseband combiner design at the BBU, while satisfying the latency requirement of the corresponding computational tasks. Compared to \cite{Barbarossa2013} and \cite{Barbarossa2015} where the latency requirement only includes the transmission latency and computational latency, we further consider the fronthaul latency experienced in transferring the quantized bits from the xL-MIMO RRH to the BBU. This additional fronthaul latency couples with the transmission latency through the required number of quantization bits and makes our non-convex optimization problem fundamentally different from the existing works  \cite{Barbarossa2013, Barbarossa2015, li2018minmax}, which is more challenging to solve.

\item To obtain a near-optimal solution for the formulated optimization problem, we introduce a two-stage design, where a hybrid spatial filtering (HSF) matrix at the xL-MIMO RRH is firstly obtained purely based on the channel state information (CSI). Subsequently, based on the effective channel and the obtained HSF matrix, a joint optimization on the residual variables at the BBU is implemented. For the joint optimization at the BBU, the proposed problem is divided into three sub-problems and solved via alternating optimization. To be more specific, the baseband combiner is obtained in a closed-form, the communication and computational resource allocation problem is solved by leveraging the  successive inner convexification, and the optimization on the number of quantization bits is solved through a line-search method. Moreover, the proposed algorithm is shown to converge to a local optimal solution. 

\item For practical implementation, we resort to deep learning  for a low-complexity solution for the joint optimization at the BBU \cite{pcDeep2018, learningSurvey2019, decode2018, DeepUSVXu2019, DeepSVKim2018}. Specifically, we deploy a supervised learning method using the DNN, where the Adam optimizer \cite{kingma2014adam} is used to train the DNN with the solutions obtained from the complicated optimization algorithm. Finally, the numerical results demonstrate the superiority of the proposed joint optimization algorithm over the disjoint optimization procedures. Furthermore, the DNN based supervised learning is shown to be an effective low-complexity approach, which reduces the execution time by two orders of magnitudes.

\end{enumerate}

The rest of the paper is organized as follows. Section II describes the system model and introduces the formulated problem for the total transmit power minimization. In Section III, we present the proposed solution for the formulated problem, where the hybrid spatial filtering matrix at the xL-MIMO RRH is obtained locally followed by the joint optimization of the residual parameters at the BBU via alternating optimization, and finally discuss the low-complexity solution for the joint optimization based on the DNN method. Numerical results are presented in Section IV, and we conclude the paper in Section V.

\textit{Notations}: Bold upper-case letters  $\mathbf{Y}$, bold lower-case letters $\mathbf{y}$ and letters $y$ denote matrices, vectors and scalars, respectively; $Y_{i,j}$ is the entry on the $i$-th row and $j$-th column of $\mathbf{Y}$; Transpose and conjugate transpose of $\mathbf{Y}$ are represented by $\mathbf{Y}^T$ and $\mathbf{Y}^H$, respectively; $\mathbf{Y}^\dagger$ is the Moore-Penrose pseudo inverse of $\mathbf{Y}$; ${\rm diag}\left(\left[y_1, \dots, y_n \right]^T\right)$ denotes a diagonal matrix with elements $y_i, \; i = 1, \dots, n$ on the diagonal; ${\rm vec}(\mathbf{Y})$ indicates vectorization; $\norm{\bf y}_2$ is the $\ell_2$ norm of the vector ${\bf y}$; $\mathbf{1}_M$ is the $M \times 1$ vector of ones;  $\jmath$ is defined as  $\jmath \triangleq \sqrt{-1}$,  $|\cdot|$ returns the amplitude of a complex number; $\odot$, $\oslash$ and $\circ$ denote the Hadamard product, division and power, respectively;  ${\bf I}$ is the identity matrix.

%\vspace{-2mm}

\section{System Model and Problem Formulation}

%\vspace{-2mm}
\begin{figure}[!htb]
       \centering
                    \includegraphics[scale=0.375]{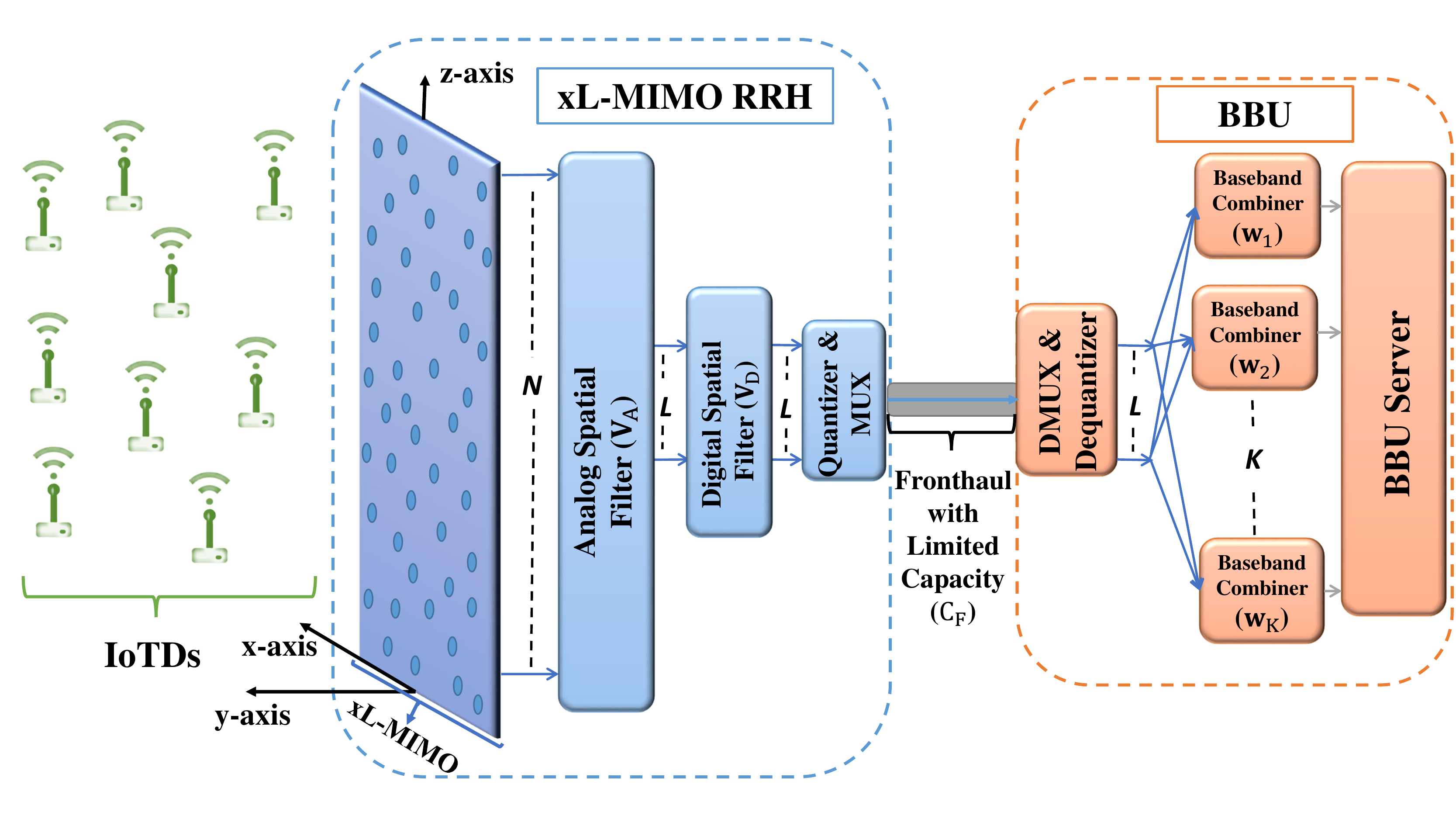}
\caption{\small System model of the uplink xL-MIMO C-RAN serving $K$ IoTDS.}
   \label{SM}
\end{figure}
\vspace{-2mm}

We consider the uplink of an xL-MIMO C-RAN that serves $K$ single-antenna IoTDs, as shown in Fig.~\ref{SM}. The xL-MIMO RRH consists of $N$ antennas uniformly distributed in a two-dimensional space along the $xz$-plane at $y = 0$ on the Cartesian coordinates. Accordingly, the locations of the $n$-th antenna of the xL-MIMO RRH and the $k$-th IoTD are defined as $\left(x_{n}^{M}, 0, z_{n}^{M} \right)$ and $\left(x_k, y_k, z_k \right)$, respectively. In this paper, we assume that the xL-MIMO RRH is equipped with $R = K \left(\ll N \right)$ RF chains such that there are enough spatial degrees of freedom to serve all the $K$ IoTDs \cite{liu2019TwoScale}. The xL-MIMO RRH is connected to the BBU\footnote{Note that the BBU can be shared among multiple xL-MIMO systems, thereby reducing the total cost of ownership (TCO) \cite{CRanSurvey2014}.}
 via a digital error-free fronthaul link with a capacity of $C_F$ bits per second (bps). The BBU makes the resource allocation decisions and  decodes  the IoTDs’ symbols, followed by the processing of the  computation bits at the BBU server. 
We assume that all the IoTDs transmit over a quasi-static flat-fading channel, and the received signal at the xL-MIMO RRH is expressed as 

\begin{equation}
    {\bf y} = \sum_{k = 1}^K {\bf h}_k \sqrt{p_k} s_k + {\bf z},
\end{equation}
where $s_k$ is the transmitted symbol of the $k$-th IoTD such that $|s_k|^2 = 1$, $p_k$  is the corresponding transmit power, ${\bf h}_k \in \mathbb{C}^{N \times 1}$ is the channel vector between the $k$-th IoTD and the xL-MIMO RRH, and ${\bf z} \sim \mathcal{CN}(0, \sigma^2 {\bf I}_N)$ denotes the additive white Gaussian noise (AWGN).

\vspace{-4mm}

\subsection{Channel Model}

Given that the IoTDs are deployed in an indoor environment, the IoTDs are reasonably close to the xL-MIMO RRH. Hence, the desired channel between an IoTD and each antenna of the xL-MIMO RRH is composed of both the deterministic LOS and non-line-of-sight (NLOS) components. Accordingly, the channel between the $k$-th IoTD and the xL-MIMO RRH is given by \cite{LISWaad2018}

%\vspace{-4mm}

\begin{equation}
    {\bf h}_k = {\bm \kappa}_k^{L} {\bf h}^L_k + {\bm \kappa}_k^{NL} {\bf h}^{NL}_k,
\end{equation}
where ${\bf h}^L_k \in \mathbb{C}^{N \times 1}$ is the deterministic LOS component between the $k$-th IoTD and the xL-MIMO RRH, given by \cite{LISWaad2018}

\vspace{-2mm}

\begin{equation}
    {\bf h}^L_k = \left[l_{1,k}^L h_{1,k} , \dots, l_{N,k}^L h_{N,k} \right]^T,
\end{equation}
where $l_{n,k}^L  = \frac{1}{\sqrt{4 \pi d_{n,k}^2}}$, $h_{n,k} = \exp\left(\frac{-j 2 \pi d_{n,k}}{\lambda} \right)$ and $d_{n,k}  = \sqrt{\left(x_k - x_{n}^{M}\right)^2 + y_{k}^2 + \left(z_k - z_{n}^{M}\right)^2}$ are the attenuation factor in the free space, the channel gain and the distance between the $k$-th IoTD and the $n$-th antenna of the xL-MIMO RRH, respectively, with $\lambda$ denoting the carrier wavelength of the transmitted signal. The NLOS component ${\bf h}^{NL}_{k} \in \mathbb{C}^{N \times 1}$ between the $k$-th IoTD and the xL-MIMO RRH is defined as \cite{DeepUSVXu2019}

\vspace{-2mm}

\begin{equation}
  {\bf h}^{NL}_k = {\bm \Lambda}^{\frac{1}{2}}_k {\bf g}_k,
\end{equation}
with ${\bm \Lambda}_{k} \triangleq {\rm diag}\left(\left[d_{1,k}^{-\xi} \tau_{1,k}, \dots, d_{N,k}^{-\xi}\tau_{N,k}  \right]^T \right) \in \mathbb{C}^{N \times N}$, where $d_{n,k}^{-\xi}$  and $\tau_{n,k}$ are the large-scale fading and the log-normal shadow fading between the $k$-th IoTD and the $n$-th antenna of the xL-MIMO RRH,  respectively. $\xi$ is the path loss exponent and ${\bf g}_k \in \mathbb{C}^{N \times 1}$ models the small-scale fading, with each entry following $\mathcal{CN} \left(0,1 \right)$. Finally, ${\bm \kappa}_k^{L} \triangleq {\rm diag}\left( \left[ \sqrt{\frac{\kappa_{1,k}}{\kappa_{1,k}+1}}, \dots, \sqrt{\frac{\kappa_{N,k}}{\kappa_{N,k}+1}} \right]^T \right) \in \mathbb{R}^{N \times N}$ and ${\bm \kappa}_k^{NL}  \triangleq {\rm diag}\left( \left[ \sqrt{\frac{1}{\kappa_{1,k}+1}}, \dots, \sqrt{\frac{1}{\kappa_{N,k}+1}} \right]^T \right) \in \mathbb{R}^{N \times N}$, where $\kappa_{n,k}$ denotes the Rician factor between the $k$-th IoTD and the $n$-th antenna of the xL-MIMO RRH.

\subsection{Uplink Signal Processing}

At the xL-MIMO RRH, we consider the spatial-compression-and-forward (SCF) scheme proposed in \cite{LiuJointPower2015,liu2019TwoScale, zhang2015CRAN} to balance between the information conveyed to the BBU and the data traffic over the fronthaul link. To reduce the hardware complexity, we employ the hybrid analog-digital filtering, where each antenna is only equipped with a phase shifter and the signals from  $N$ antennas are filtered using an analog spatial filtering matrix ${\bf V}_{A} \in \mathbb{C}^{N \times R}$, followed by a digital spatial filtering matrix ${\bf V}_{D} \in \mathbb{C}^{R \times R}$. Accordingly, the received signal after the hybrid spatial filtering is given by

\begin{equation}
 \begin{split}
    \bar{\bf y} &= {\bf V} {\bf y}, \\
    &= {\bf V} \sum_{k = 1}^K {\bf h}_k \sqrt{p_k} s_k + {\bf V} {\bf z},
 \end{split}
\end{equation}
where ${\bf V} \triangleq {\bf V}_D^H {\bf V}_A^H = \left[{\bf v}_1, \dots, {\bf v}_R \right]^T \in \mathbb{C}^{R \times N}$ denotes the HSF matrix. With the use of phase shifters, each entry of $\mathbf{V}_{A}$ satisfies the element-wise constant-modulus constraint, i.e.,  $\left|\left[{\bf V}_A \right]_{\left(i,j \right)} \right|= 1, \; \forall i,j$. In this paper, we assume that high-resolution ADCs are used at the xL-MIMO RRH such that the quantization error due to ADCs is negligible \cite{liu2019TwoScale}. Subsequently, a uniform scalar quantization is applied to each element of $\bar{{\bf y}} = [\bar{y}_{1}, \dots, \bar{y}_{R}]^T$, where each complex symbol $\bar{y}_{r}$ can be represented by its in-phase (I) and quadrature (Q) part as

%\vspace{-2mm}

\begin{equation}
  \bar{y}_{r} = \bar{y}^{I_r} + \jmath\;  \bar{y}^Q_{r}, \; \forall r,   
\end{equation}
where the I-branch symbol $\bar{y}^I_{r}$ and Q-branch symbol $\bar{y}^Q_{r}$  are both real Gaussian random variables with zero mean and variance $\left(\sum_{k = 1}^K p_k |{\bf v}_r^T {\bf h}_{k}|^2 + \sigma^2 \norm{{\bf v}_r}^2 \right)/2$ \cite{LiuJointPower2015}. After the uniform scalar quantization, the baseband quantized symbol of $\bar{\bf y}$ is given by

%\vspace{-2mm}

\begin{equation}
\begin{split}
        \tilde{\bf y} &= \bar{\bf y} + {\bf e}, \\ &=  {\bf V} \sum_{k = 1}^{K}  {\bf h}_k \sqrt{p_k} s_k + {\bf V z} + {\bf e},
\end{split}
\end{equation}
where ${\bf e} \triangleq \left[e_{1}, \dots, e_{R} \right]^T$ denotes the additive quantization error vector for  $\bar{\bf y}$. Each ${e}_r$ is Gaussian distributed with zero mean and variance $\varrho_r$, with $\varrho_r$ given by  \cite{LiuJointPower2015}

\begin{equation}
  \varrho_r=\begin{cases}
    3 \left(\sum_{k=1}^K p_k \left|{\bf v}^T_{r} {\bf h}_{k} \right|^2 + \sigma^2 ||{\bf v}_{r}||^2 \right) 2^{-2 \varpi}, & \text{if  $\varpi > 0$},\\
    \infty, & \text{if $\varpi = 0$},
  \end{cases}
\end{equation}
%
%\begin{equation}
%  \varrho_r=  3 \left(\sum_{k=1}^K p_k \left|{\bf v}^T_{r} {\bf h}_{k} \right|^2 + \sigma^2 %||{\bf v}_{r}||^2 \right) 2^{-2 \varpi},
%\end{equation}
%
where $\varpi$ denotes the number of bits that each RF chain uses to quantize $\bar{y}^I_{r}$ and $\bar{y}^Q_{r}$.  As each $e_r$ is independent over $r$ due to the independent scalar quantization for each element of $\bar{\bf y}$, and therefore the covariance matrix of ${\bf e}$ is a function of ${\bf p} \triangleq \left[p_1, \dots, p_K \right]^T$, ${\bf V}$ and $\varpi$, given by

\begin{equation}
    {\bf Q} ({\bf p}, {\bf V}, {\varpi}) = \mathbb{E}[{\bf e} {\bf e}^H] = {\rm diag}\left(\left[\varrho_{1}, \dots, \varrho_r\right]^T\right).
\end{equation}

Subsequently, the quantized symbols are forwarded to the BBU via the fronthaul link. To mitigate the effects of the inter-IoTD interference and the quantization error, a linear baseband combiner ${\bf w}_k \triangleq \left[ w_{k,1}, \dots, w_{K,R}\right]^T \in \mathbb{C}^{R \times 1}$ is further applied to $\tilde{\bf y}$ before demodulating the symbol for the $k$-th IoTD, given by

%\vspace{-2mm}

\begin{equation}
\begin{split}
    \hat{s}_k &= {\bf w}^H_k \tilde{\bf y}, \\
    &= {\bf w}^H_k {\bf V} {\bf h}_k \sqrt{p_k} s_k + \sum_{\substack{j = 1, \;j \neq k}}^K {\bf w}^H_k {\bf V} {\bf h}_j \sqrt{p_j} s_j + {\bf w}^H_k {\bf V} {\bf z} + {\bf w}^H_k {\bf e}.
\end{split}
\end{equation}
Accordingly, the SINR for the $k$-th IoTD is expressed as

\begin{equation}
    \gamma_k\left({\bf p}, {\bf V}, {\bf W}, \varpi \right) = \frac{p_k \left| {\bf w}^H_k {\bf V} {\bf h}_k \right|^2}{\sum_{\substack{j = 1, \;j \neq k}}^K p_j \left|{\bf w}^H_k {\bf V} {\bf h}_j \right|^2 + \sigma^2 \norm{{\bf w}^H_k {\bf V}}^2 + {\bf w}^H_k {\bf Q}({\bf p}, {\bf V}, {\varpi}) {\bf w}_k},
\end{equation}
where  ${\bf W} \triangleq \left[{\bf w}_1, \dots, {\bf w}_K \right]$.

\subsection{Computation Offloading and Latency Model}

We assume that due to the limited computational capability at the IoTDs, all the computational tasks of the IoTDs have to be offloaded to the BBU. Accordingly, let the $k$-th IoTD's computational task $C_k$ be described by a tuple, defined as  $\left(\omega_k,b_k,\mathcal{T}^{th}_k \right)$, where $\omega_k$ denotes the number of CPU cycles needed for computing $C_k$, $b_k$ represents the number of computation bits needed for $C_k$ and $\mathcal{T}^{th}_k$ is the maximum tolerable latency to execute $C_k$ \cite{Barbarossa2015}. In the case of offloading, the latency includes a) the transmission latency, b) the fronthaul latency,  and c) the computational latency.

\subsubsection{Transmission latency $(\xi_{k}^{TL})$}

Given $\gamma_k\left({\bf p}, {\bf V}, {\bf W}, \varpi \right),\;\forall k$, the transmission latency $\xi_k^{TL}$ is incurred during the transmission of the computation bits $b_k$ from the $k$-th IoTD to the xL-MIMO RRH. The latency for the transmission of $\log_2\left(1 + \gamma_k \left({\bf p}, {\bf V}, {\bf W}, \varpi\right)\right)$ bits per second per Hertz is given by \cite{li2018minmax, Barbarossa2013, Barbarossa2015}

\begin{equation}
    \xi^{TL}_k = \frac{b_k}{B_W \log_2\left(1 + \gamma_k\left({\bf p}, {\bf V}, {\bf W}, \varpi \right)\right)},
\end{equation}
where $B_W$ is the total transmission bandwidth.

%\vspace{-4mm}

\subsubsection{Fronthaul latency $(\xi_k^{FL})$}

For $b_k$ computation bits corresponding to the $k$-th IoTD, we assume that the bits are encoded using the $M$-PSK modulation. Accordingly, $b_k$ bits are encoded into $\frac{b_k}{\log_2(M)}$ symbols which are transmitted from the $k$-th IoTD to the BBU through the xL-MIMO RRH.  With $\varpi$ bits used to quantize both the real and imaginary part of each entry in $\bar{\bf y}$, a total of  $2 R {\varpi}$ quantized bits are required across $R$ RF chains \cite{zhang2015CRAN, LiuJointPower2015}. Consequently, $\frac{b_k}{\log_2(M)}$ transmitted symbols of the $k$-th IoTD generate effective traffic of $\frac{2 b_k R \varpi}{\log_2(M)}$ bits for the fronthaul link. Hence, with a fronthaul link capacity of $C_F$, expressed in terms of bits per second, the fronthaul latency $\xi_k^{FL}$ for forwarding  $b_k$ computation bits of the $k$-th IoTD from the xL-MIMO RRH to the BBU is given  by \cite{li2018minmax}

\begin{equation}
    \xi^{FL}_k = \frac{2 b_k R \varpi}{C_F \log_2(M)}.
\end{equation}

\subsubsection{Computational latency $(\xi_k^{CL})$}

 The computational resources are shared among the $K$ IoTDs and are quantified by the computational rate $F_T$, expressed in terms of the number of CPU cycles per second \cite{Barbarossa2015, li2018minmax}. Let us denote by $f_k \geq 0$ the fraction of $F_T$ to be assigned to each IoTD. The rates $f_k$ are subject to the computational budget constraint, i.e., 

\begin{equation}
   \sum_{k=1}^K f_k \leq F_T.   
\end{equation}
Given the resource assignment $f_k$, the computational latency $\xi_k^{CL}$ incurred in executing $\omega_k$ CPU cycles for the computational task of the $k$-th IoTD is given by \cite{li2018minmax, Barbarossa2013, Barbarossa2015}

\begin{equation}
   \xi_k^{CL} = \frac{\omega_k}{f_k}, \forall k.     
\end{equation}

Finally, the expression for the overall latency $\xi_k$ is given by

\begin{equation}
\begin{split}
   \xi_k &= \xi_{k}^{TL} + \xi^{FL}_k +  \xi_{k}^{CL}, \\
   &= \frac{b_k}{B_W \log_2\left(1 + \gamma_k \left({\bf p}, {\bf V},  {\bf W}, \varpi\right)\right)} +\frac{2 b_k R \varpi}{C_F \log_2(M)} + \frac{\omega_{k}}{f_k}. 
   \end{split}
\end{equation}
{(16) clearly shows the interplay between the wireless transmission part and the computational part via the transmission and computational latency. Furthermore, a coupling between the transmission and fronthaul latency through the number of quantization bits $\varpi$ can also be observed from (16). For example, an increase in the quantization bits decreases the quantization error which reduces the transmission latency, while on the other hand, it increases the required number of bits transmitted to the BBU, thereby increasing the fronthaul latency. Therefore, the joint optimization of the communication and computational resource allocations along with the number of quantization bits for the computation offloading task is essential.

\subsection{Problem Formulation}

In this paper, we aim to minimize the total transmit power for the IoTDs, i.e., ${\bf 1}_K^T {\bf p}$, by jointly optimizing the communication resource ${\bf p}$, the HSF matrix ${\bf V}$, the baseband combiner ${\bf W}$ at the BBU, the number of quantization bits ${\varpi}$ and the computational resource ${\bf f} \triangleq \left[f_1, \dots, f_K \right]^T$.  Accordingly, we aim to solve the following optimization problem:

\begin{equation}
\begin{aligned}
& \mathcal{P}_1: && \underset{\left\{{\bf V}, {\bf W}, {\bf p}, {\bf f}, {\varpi}\right\}}{\min} \;{\bf 1}_K^T {\bf p} \\ % E_k \left(p_k, {\bf V}, {\bf w}_k, {\bf D} \right)    \\
& \text{\it s.t.}
& & {\rm C_1}:\frac{b_k}{B_W \log_2\left(1 + \gamma_k\left({\bf p}, {\bf V}, {\bf W}, \varpi \right)\right)} +\frac{2 b_k R \varpi}{C_F \log_2(M)} + \frac{\omega_{k}}{f_k} \leq  \mathcal{T}_k^{th}, \; \forall k,\\
%& & {\rm C_1}: \xi_{k} \leq  \mathcal{T}_k^{th}, \; \forall k \\
&&& {\rm C_2}: {\bf 1}^T_K {\bf f}  \leq F_T, \;{\rm C_3}: p_k \leq P_{k,max}, \; \forall k, \\
&&& {\rm C_4}: 2 B_W R {\varpi} \leq C_{F}, \; {\rm C_5}: {\varpi} \in \mathbb{Z}_{>0},\\
&&& {\rm C_6}: \left|\left[{\bf V}_A \right]_{\left(i,j \right)} \right|= 1, \; \forall i,j
\end{aligned}
\end{equation}
where ${\rm C_1}$ is the latency constraint for the $k$-th IoTD with the latency threshold denoted by $\mathcal{T}_k^{th}$,  ${\rm C_2}$ is the computational resource constraint, ${\rm C_3}$ is the maximum power limits of the IoTDs, where $P_{k,max}$ denotes the maximum transmit power of each IoTD, ${\rm C_4}$ is the fronthaul capacity constraint \cite{Barbarossa2015, zhang2015CRAN, liu2019TwoScale}, ${\rm C_5}$ is the integer constraint for the number of quantization bits, and ${\rm C_6}$ is the element-wise constant-modulus constraint for the analog spatial filtering matrix.

\section{{Proposed solution for the Formulated Problem $\mathcal{P}_1$}}

In this section, we seek a feasible solution for $\mathcal{P}_1$, which is found to be non-convex due to 1) the coupling of  variables  between  the  transmission latency, the fronthaul latency and the computational latency, 2) the integer constraint for the quantization bit, and 3) the element-wise constant-modulus constraint  for  the  analog  spatial  filtering  matrix.  Accordingly, to solve $\mathcal{P}_1$, we present a two-stage design, where the HSF matrix at the xL-MIMO RRH is obtained locally\footnote{Designing the HSF matrix locally at the xL-MIMO RRH reduces the signaling overhead between the the BBU and the xL-MIMO RRH.} based on the CSI, and a joint optimization on the residual variables at the BBU is subsequently implemented based on the effective channel and the obtained HSF matrix. Finally, we propose a low-complexity solution based on the deep learning framework to address the practical implementation of the considered joint optimization at the BBU.

  \vspace{-4mm}

\subsection{HSF Design at the xL-MIMO RRH}

In this work, the HSF matrix is obtained by approximating the fully-digital spatial filtering  (FDSF) matrix. To pursue a low-complexity solution, we select the matched filtering (MF) method as the FDSF, given by
 
 %\vspace{-2mm}

 \begin{equation}
     {\bf V}_{FD} = {\bf H}^H,
 \end{equation}
where ${\bf H} \triangleq \left[{\bf h}_1, \dots, {\bf h}_K \right]^T$. Another advantage of employing the MF approach is that ${\bf V}_{FD}$ tends to eliminate the effect of small-scale fading, resulting in a less frequent update of the communication and computational parameters at the BBU. Specifically, assuming the independence of each individual propagation path of the IoTDs and by leveraging the concept of channel hardening and law of large numbers \cite{ngo2013energy, LISWaad2018}, it can be shown that the effective channel at the BBU, given by ${\bf V}_{FD} {\bf H}$, tends to be independent of the small-scale fading, i.e.,

%\vspace{-2mm} %with a low implementation complexity

\begin{equation}
\begin{split}
    {\bf V}_{FD} {\bf H} &\xrightarrow[]{N \to \infty} {\rm diag}\left(\left[{ h}_{eff}^{1}, \dots, {h}_{eff}^{K} \right]^T\right),
    \end{split}
\end{equation}
where ${h}_{eff}^{k} \triangleq \sum_{n = 1}^N {\frac{\kappa_{n,k}}{\kappa_{n,k}+1}} \left|l_{n,k}^L h_{n,k} \right|^2 + {\frac{1}{\kappa_{n,k}+1} \left|d_{n,k}^{-\xi}\right|}, \; \forall k$.  Accordingly, ${\bf V}_{FD}$ asymptotically decorrelates the signals from the IoTDs across the $K$ output dimensions. Subsequently, for a given FDSF matrix, the hybrid analog and digital spatial filtering matrix, i.e., $ {\bf V}_A$ and  ${\bf V}_D$, are obtained by \cite{SVDlow2018}

\vspace{-2mm}

\begin{equation}
    {\bf V}_A = {\bf V}_{FD} \oslash \left|{\bf V}_{FD} \right|,
\end{equation}
and 
\begin{equation}
    \mathbf{{V}}_{D} = {\bf V}_A^{\dagger} {\bf V}_{FD}.
\end{equation}
Consequently, the HSF matrix is given by ${\bf V} = {\bf V}_D^H {\bf V}_A^H$.

% denoted by  ${\bf x} \triangleq \left[\left({\rm vec}\left({\bf W}\right)\right)^T, {\bf p}^T, {\bm \varpi}^T, {\bf f}^T\right]^T$
\vspace{-2mm}

\subsection{Joint Optimization at the BBU}

Next, we propose to solve a joint optimization on the residual variables at the BBU based on the alternating optimization framework\footnote{Alternating optimization  has been extensively used in applications such as  image processing \cite{wang2008new}, robust learning \cite{jain2017non}, wireless signal processing \cite{zhang2015CRAN,liu2019TwoScale}, etc, where the optimization problem concerning two or more variables are solved by  fixing one or group of the variables and optimizing over the others \cite{bezdek2002some, bezdek2003convergence, jain2013low}.}, which effectively removes the coupling between the transmission latency, the fronthaul latency and the computational latency. To be more specific, given the HSF matrix ${\bf V}$, $\mathcal{P}_1$ can be transformed into a joint optimization on ${\bf W}$, ${\bf p}$, ${\bf f}$ and ${\varpi}$, given by
 \vspace{-2mm}

\begin{equation}
\begin{aligned}
& \mathcal{P}_2: && \underset{
\left\{{\bf W}, {\bf p}, {\bf f}, {\varpi}\right\}}{\min} \;  {\bf 1}_K^T {\bf p}\\ % E_k \left(p_k, {\bf w}_k, {\bf D} \right)    \\
& \text{\it s.t.}
& & {\rm C_1}: \frac{b_k}{B_W \log_2\left(1 +\gamma_k\left({\bf p}, {\bf W}, \varpi \right)\right)} +\frac{2 b_k R \varpi}{C_F \log_2(M)} + \frac{\omega_{k}}{f_k} \leq  \mathcal{T}_k^{th}, \; \forall k\\
&&& {\rm C_2}: {\bf 1}^T_K {\bf f}  \leq F_T, \; {\rm C_3}: p_k \leq P_{k,max}, \; \forall k, \\
&&& {\rm C_4}: 2 B_W R {\varpi} \leq C_{F}, \; {\rm C_5}: {\varpi} \in \mathbb{Z}_{>0}.
\end{aligned}
\end{equation}
It should be noted that the HSF matrix in (21) does not completely decorrelate the signals from the IoTDs due to the finite number of antennas at the xL-MIMO RRH, resulting in the inter-IoTD interference. This along with the quantization noise introduced by the subsequent quantizer may degrade the demodulation performance of the signals at the BBU. Hence, we further adopt a baseband combiner at the BBU to obtain the received symbols as close as possible to the original symbols. Consequently, following the  minimum-mean-squared-error (MMSE)  metric and for a given ${\bf p}$, ${\bf f}$ and $\varpi$, the optimal linear baseband combiner for $\mathcal{P}_2$ is given by \cite{zhang2015CRAN, liu2019TwoScale}

\begin{equation}
    \bar{\bf w}_k = \left( \sum_{j = 1}^K p_j \left|{\bf V} {\bf h}_j \right|^2 + \sigma^2 {\bf I} + {\bf Q}({\bf p}, {\bf V}, {\varpi})  \right)^{-1} {\bf V} {\bf h}_k.
 \end{equation}
Based on the fact that

\begin{equation}
    \bar{\bf w}_k^H {\bf Q}({\bf p}, {\bf V}, {\varpi}) \bar{\bf w}_k = \sum_{r = 1}^R \varrho_{r} \left|\bar{w}_{k,r} \right|^2 =  2^{-2 \varpi} \sum_{r = 1}^R \Xi_{r,k} \left({\bf p}\right),
\end{equation}
where  $\bar{w}_{i,j}$ denotes the $j$-th element  of $\bar{\bf w}_i$, $1 \leq i \leq K$, $1 \leq j \leq R$ and 

\begin{equation}
    \Xi_{r,k} \left({\bf p} \right) = 3 \left|\bar{w}_{k,r} \right|^2 \left(\sum_{j=1}^K p_j \left|{\bf v}^T_{r} {\bf h}_{j} \right|^2 + \sigma^2 ||{\bf v}_{r}||^2 \right), \; \forall l,
\end{equation}
and by defining $\alpha_{k,j} \triangleq \left| \bar{\bf w}^H_k {\bf V} {\bf h}_j \right|^2$, $\eta_k \triangleq \sigma^2 \norm{\bar{\bf w}^H_k {\bf V}}^2$, (12) can be expressed as a function of ${\bf p}$ and ${\varpi}$ as

\begin{equation}
\begin{split}
     \xi_k^{TL} \left({\bf p}, {\varpi} \right) &\triangleq \frac{b_k}{B_W \log_2 \left(1 + \frac{p_k \alpha_{k,k}}{\eta_k +  \sum_{\substack{j = 1, \;j \neq k}}^K p_j  \alpha_{k,j} + 2^{-2 \varpi} \sum_{l=1}^R \Xi_{l,k} \left({\bf p} \right)} \right)}. 
\end{split}
\end{equation}
% \frac{b_k}{B_W \log_2\left(1 + \gamma_k\right)}, \\
Accordingly, $\mathcal{P}_2$ is transformed into a joint optimization on ${\bf p}$,  ${\bf f}$, and ${\varpi}$, given by

\begin{equation}
\begin{aligned}
& \mathcal{P}_3: && \underset{\left\{{\bf p}, {\bf f}, {\varpi}\right\}}{\min} \; {\bf 1}_K^T {\bf p}\\ % E_k \left(p_k, {\bf w}_k, {\bf D} \right)    \\
& \text{\it s.t.}
& & {\rm C_1}:  \xi_k^{TL} \left({\bf p}, {\varpi} \right)  +\frac{2 b_k R \varpi}{C_F \log_2(M)} + \frac{\omega_{k}}{f_k} \leq  \mathcal{T}_k^{th}, \; \forall k\\
&&& {\rm C_2}: {\bf 1}^T_K {\bf f}  \leq F_T, \; {\rm C_3}: p_k \leq P_{k,max}, \; \forall k, \\
&&& {\rm C_4}: 2 B_W R {\varpi} \leq C_{F}, \; {\rm C_5}: {\varpi} \in \mathbb{Z}_{>0}.
\end{aligned}
\end{equation}
Based on the formulation, we discuss the feasibility of $\mathcal{P}_3$, as shown in Lemma 1 below.

\noindent
\textbf{\textit{Lemma 1:}} $\mathcal{P}_3$ admits a non-empty feasible set satisfying all the constraints in (27), if for $\mathcal{T}^{th}_k > 0, \; \forall k$, $\exists$ ${\bf p} \in \Psi \triangleq \left\{\bar{\bf {p}} \in \mathbb{R}^K_+ : \bar{\bf {p}} \preccurlyeq {\bf P}_{max} \right \}$ and ${{\varpi}} \in \mathcal{D} \triangleq \left\{\bar{\varpi} \in \mathbb{Z}_{>0}: \bar{\varpi} \leq \frac{C_{F}}{2 R B_W} \right\}$, where ${\bf P}_{max} \triangleq \left[P_{1,max}, \dots, P_{K,max} \right]^T$, the following sufficient and necessary conditions are satisfied:

\begin{subequations}
    \begin{equation}
        \xi_k^{TL} \left({\bf p}, {\varpi} \right) + \frac{2 b_k R  {\varpi}}{C_F \log_2(M)} < \mathcal{T}^{th}_k, \; \forall k,
    \end{equation}
    \begin{equation}
             \sum_{k = 1}^K \frac{\omega_k}{{\mathcal{T}}_k^{th} - \xi_k^{TL} \left({\bf p}, {\varpi} \right) - \frac{2 B_W R \varpi}{C_F} } \leq F_T.
    \end{equation}
\end{subequations}

 \textit{Proof}: The individual conditions in (28a) are necessary to ensure that each IoTD can transmit the computation bits to the BBU within the maximum tolerable latency. Subsequently, (28b) guarantees that the total computational resource available at the BBU is enough to assign  the computational resource to each IoTD to execute their computational tasks while satisfying the corresponding latency requirement. $\blacksquare$

In what follows, we assume that $\mathcal{P}_3$ is feasible\footnote{The conditions in (28) can be enforced by a proper admission control strategy \cite{Barbarossa2013, addOffload2018}, or an appropriate choice of the fronthaul capacity or the BBU computational capability \cite{offStrategy2019}.} and present the corresponding solution. Accordingly, to solve $\mathcal{P}_3$, we first fix the number of  quantization bits $\varpi$ in $\mathcal{P}_3$ and optimize ${\bf p}$ and ${\bf f}$ by solving the following sub-problem:

\begin{equation}
\begin{aligned}
& \mathcal{P}_4: && \underset{\left\{{\bf p}, {\bf f}\right\}}{\min} \; {\bf 1}_K^T {\bf p} \\ % E_k \left(p_k, {\bf w}_k, {\bf D} \right)    \\
& \text{\it s.t.}
& & {\rm C_1}: \frac{b_k}{B_W \log_2 \left(1 + \frac{p_k \alpha_{k,k}}{\eta_k +  \sum_{\substack{j = 1, \;j \neq k}}^K p_j  \alpha_{k,j} + 2^{-2 \varpi} \sum_{l=1}^R \Xi_{l,k} \left({\bf p} \right)} \right)} + \frac{\omega_{k}}{f_k} \leq  \bar{\mathcal{T}}_k^{th}, \; \forall k\\
&&& {\rm C_2}: {\bf 1}^T_K {\bf f}  \leq F_T,  \; {\rm C_3}: p_k \leq P_{k,max}, \; \forall k, \\
\end{aligned}
\end{equation}
where $\bar{\mathcal{T}}_k^{th} \triangleq \mathcal{T}_k^{th} -\frac{2 b_k R \varpi}{C_F \log_2(M)}$. Subsequently, the number of quantized bits ${\varpi}$ is obtained through the following feasibility problem

\begin{equation}
\begin{aligned}
& \mathcal{P}_{5}: && {\text {Find}} \; \{\varpi\}  \\
& \text{\it s.t.}
& &  {\rm C_1}:\frac{b_k}{B_W \log_2 \left(1 +  \frac{p_k \alpha_{k,k}}{ \tilde{\eta}_k +  2^{-2 \varpi} \sum_{r=1}^R \Xi_{r,k} \left({\bf p}\right)} \right)}  +  \frac{2 b_k R \varpi}{C_F \log_2(M)} &\leq  \tilde{\mathcal{T}}_k^{th}, \; \forall k, \\
&&& {\rm C_2}: 2 B_W R \varpi \leq C_{F}, \; {\rm C_3}: \varpi \in \mathbb{Z}_{>0}.
\end{aligned}
\end{equation}
where $\tilde{\mathcal{T}}^{th}_k \triangleq \mathcal{T}^{th}_k - \frac{\omega_{k}}{f_k}$ and $\tilde{\eta}_k \triangleq \eta_k + \sum_{\substack{j = 1, \;j \neq k}}^K p_j  \alpha_{k,j}$. 

%\vspace{-2mm}

\subsubsection{Solution for the problem $\mathcal{P}_4$} 

$\mathcal{P}_4$ is still non-convex and difficult to solve due to its constraint ${\rm C}_1$, which can be expressed as

\begin{equation}
    \begin{split}
      &\underbrace{\frac{b_k}{B_W \log_2 \left(1 +  \frac{p_k \alpha_{k,k}}{\eta_k +  \sum_{\substack{j = 1, \;j \neq k}}^K p_j  \alpha_{k,j} + 2^{-2 \varpi} \sum_{r=1}^R \Xi_{r,k}\left({\bf p} \right)} \right)} +\frac{\omega_{k}}{f_k} - \bar{\mathcal{T}}_k^{th} }_{g_k\left({\bf p}, f_k \right)}  \leq  0, \\
       \Rightarrow &\underbrace{- \log_2 \left(1 +  \frac{p_k \alpha_{k,k}}{\eta_k +  \sum_{\substack{j = 1, \;j \neq k}}^K p_j  \alpha_{k,j} + 2^{-2 \varpi} \sum_{r=1}^R \Xi_{r,k}\left({\bf p} \right)} \right)}_{g^{'}_k\left({\bf p}\right): \; \text {non-convex}} + \underbrace{\frac{f_k b_k}{B_W  f_k \bar{\mathcal{T}}_k^{th} - \omega_k}}_{g^{'}_k\left({f_k}\right): \;\text{convex}} \leq 0,\\
    \end{split}
\end{equation}
such that $g_k\left({\bf p}, f_k\right) = g^{'}_k\left({\bf p}\right) + g^{'}_k\left({f_k}\right)$. ${\rm C}_1$ is non-convex due to $g_k\left({\bf p}, f_k\right)$. To overcome this difficulty, we exploit the framework of successive inner convexification for  $g_k({\bf p}, f_k)$ \cite{marks1978general}. The successive inner convexification optimizes a sequence of approximate convex problems, denoted by ${\cal A}_{CP}$, which allows the development of a  computationally-efficient algorithm  converging  to  a  first-order optimal solution \cite{marks1978general, zappone2016EE}. As the non-convexity of $g_k({\bf p}, f_k)$ stems from $g^{'}_k({\bf p})$, in the following we obtain a convex approximation for $g^{'}_k({\bf p})$. To be more specific, letting $p_k = 2^{q_k}$, we have

 \begin{equation}
 \begin{split}
          g^{'}_k({\bf q}) &= - \log_2 \left(1 +  \frac{2^{q_k} \alpha_{k,k}}{\eta_k +  \sum_{\substack{j = 1, \;j \neq k}}^K 2^{q_j}  \alpha_{k,j} + 2^{-2 \varpi} \sum_{r=1}^R \Xi_{r,k}\left({\bf q} \right)} \right), \\
 \end{split}
 \end{equation}
where ${\bf q} \triangleq \left[q_1, \dots, q_K \right]^T$. In the $t$-th sequence of convexification, denoted by $\tilde{g}^{'}_k({\bf q}^{(t)}; {\bf q}^{(t-1)})$, we require the following three properties to be satisfied for the convex approximation of $g^{'}_k({\bf q}^{(t)})$ \cite{marks1978general}:
 \vspace{-4mm}

\begin{subequations}
  \begin{equation}
     g_k^{'} \left({\bf q}^{(t)} \right) \leq \tilde{g}^{'}_k \left({\bf q}^{(t)}; {\bf q}^{(t - 1)}\right), \; \forall t, k,
\end{equation}
\begin{equation}
     g_k^{'} \left({\bf q}^{(t - 1)} \right) = \tilde{g}^{'}_k \left({\bf q}^{(t - 1)}; {\bf q}^{(t - 1)}\right),\; \forall k, 
\end{equation}
\begin{equation}
    \nabla g_k^{'} \left({\bf q}^{(t - 1)}\right) = \nabla \tilde{g}^{'}_k \left({\bf q}^{(t - 1)}; {\bf q}^{(t - 1)}\right),\; \forall k,
\end{equation}  
\end{subequations}
where ${\bf q}^{(t-1)}$ is the optimal solution for ${\cal A}_{CP}^{(t-1)}$. The central step of this approach is to find a suitable approximation for ${g}^{'}_k\left({\bf q}^{(t)} \right), \; \forall k$, which fulfills the requirements in (33), given by the following lemma.

\noindent
\textbf{\textit{Lemma 2:}} For a given ${\bf q}^{(t-1)} \succeq 0$, a $\tilde{g}^{'}_k \left({\bf q}^{(t)}; {\bf q}^{(t-1)} \right)$ that satisfies (33) can be defined as

\begin{equation}
\begin{split}
     \tilde{g}^{'}_k \left({\bf q}^{(t)}; {\bf q}^{(t-1)} \right) \triangleq - \psi_k^{(t-1)} \left(\log_2 (\alpha_{k,k}) + q_k^{(t)} - \log_2 \left(\bar{\eta}_k \left({\bf q}^{(t)} \right) +  \sum_{\substack{j = 1, \;j \neq k}}^K \alpha_{k,j} 2^{q_j^{(t)}}   \right) \right) - \beta_k^{(t-1)},
\end{split}
\end{equation}
where %\vspace{-2mm}

\begin{equation}
    \psi_k^{(t-1)} \triangleq \frac{\zeta_k^{(t-1)}}{1 + \zeta_k^{(t-1)}},\; \beta_k^{(t-1)} \triangleq \log_2 \left(1 + \zeta_k^{(t-1)} \right) - \frac{\zeta_k^{(t-1)}}{1 + \zeta_k^{(t-1)}} \log_2 \left(\zeta_k^{(t-1)} \right),
\end{equation}

\begin{equation}
    \zeta_k^{(t-1)} \triangleq \frac{2^{q_k^{(t-1)}} \alpha_{k,k}}{\bar{\eta}_k\left({\bf q}^{(t-1)}\right) + \sum_{{\substack{j = 1, \;j \neq k}}}^K 2^{q_j^{(t-1)}} \alpha_{k,j}},
\end{equation}
and
\begin{equation}
    \bar{\eta}_k \left({\bf q}^{(t-1)} \right) \triangleq \eta_k + 2^{-2 \varpi} \sum_{r=1}^R \Xi_{r,k}\left({\bf q}^{(t-1)} \right).
\end{equation}

 \textit{Proof}: From (32), we have 
 
 \begin{equation}
 \begin{split}
          g^{'}_k({\bf q}^{(t)}) &\overset{(a)}\leq - \psi_k^{(t-1)} \left(\log_2 (\alpha_{k,k}) + q_k^{(t)} - \log_2 \left(\bar{\eta}_k \left({\bf q}^{(t)} \right) + \sum_{{\substack{j = 1, \;j \neq k}}}^K  \alpha_{k,j} 2^{q_j^{(t)}} \right) \right) - \beta_k^{(t-1)}, \\
          &=  \tilde{g}^{'}_k \left({\bf q}^{(t)}; {\bf q}^{(t-1)} \right), \; \forall k,
 \end{split}
 \end{equation}
 where step (a) is obtained by leveraging the lower-bound of the logarithmic function \cite{zappone2016EE}, i.e., $  \log_2 \left( 1 + \zeta \right) \geq \psi \log_2 \left(\zeta\right) + \beta$, where   $\psi = \frac{\bar{\zeta}}{1 + \bar{\zeta}}$ and $\beta = \log_2 \left(1 + \bar{\zeta} \right) - \frac{\bar{\zeta}}{1 + \bar{\zeta}} \log_2 \left(\bar{\zeta} \right)$. Hence, $\tilde{g}^{'}_k \left({\bf q}^{(t)}; {\bf q}^{(t-1)} \right), \; \forall k$ satisfies (33a), where (33b) and (33c) hold at ${\bf q}^{(t)} = {\bf q}^{(t-1)}$. $\blacksquare$

Accordingly from (31), for the $t$-th sequence, we have 

\begin{equation}
\begin{split}
     g_k\left({\bf q}^{(t)}, f_k^{(t)}\right) &\leq \tilde{g}^{'}_k \left({\bf q}^{(t)}; {\bf q}^{(t-1)}\right) + g^{'}_k\left({f_k^{(t)}} \right), \\
    &= - \psi_k^{(t-1)} \left(\Gamma_k + q_k^{(t)} \right) - \beta_k^{(t-1)}  +  \frac{f_k^{(t)} b_k}{B_W  f_k^{(t)} \bar{T}_k^{th} - \omega_k} \leq 0,
\end{split}
\end{equation}
where $\Gamma_k \triangleq \left(\log_2 (\alpha_{k,k}) - \log_2 \left(\bar{\eta}_k \left({\bf q}^{(t)} \right) + \sum_{{\substack{j = 1, \;j \neq k}}}^K  \alpha_{k,j} 2^{q_j^{(t)}} \right) \right) $. As the logarithm of the sum of the exponentials is a convex function \cite{boyd2004convex}, $g_k\left({\bf q}^{(t)}, f_k^{(t)}\right), \; \forall k$ is jointly convex in ${\bf q}^{(t)}$ and ${f}_k^{(t)}$. Consequently, ignoring the sequence index $t$, the approximate convex problem ${\cal A}_{CP}^{(t)}$ for the non-convex problem $\mathcal{P}_{4}$ is given by
%\vspace{-1.75mm}

\begin{equation}
\begin{aligned}
& \mathcal{P}_{6}: && \underset{\left\{{\bf q}, {\bf f}\right\}}{\min} \;  {\bf 1}_K^T 2^{\circ {\bf q}} \\ % E_k \left(p_k, {\bf V}, {\bf w}_k, {\bf D} \right)    \\
& \text{\it s.t.}
& & {\rm C_1}:  - \psi_k \left(\Gamma_k + q_k \right) - \beta_k  +  \frac{f_k b_k}{B_W  f_k \bar{T}_k^{th} - \omega_k}  \leq 0, \; \forall k, \\
&&& {\rm C_2}: {\bf 1}^T_K {\bf f}  \leq F_T, \; {\rm C_3}: 2^{q_k} \leq P_{k,max}, \; \forall k. \\
\end{aligned}
\end{equation}
%
%\vspace{-4mm}
where $2^{\circ {\bf q}} \triangleq \left[2^{{q}_1}, \dots, 2^{{q}_K}  \right]^T$. Further, assuming ${\bf q}^{(t)} \preccurlyeq {\bf q}^{(t-1)}$\footnote{The subsequent derivations in Lemma 3 and Lemma 4 comply with this assumption.} and $\exists \; {\bf q}^{(t)}$ such that (28) is satisfied, i.e., $2^{q_k^{(t)}} \leq P_{k,max}, \; \forall t, k$, we formulate the following problem based on $\mathcal{P}_6$:

\begin{equation}
\begin{aligned}
& \mathcal{P}_{7}: && \underset{\left\{{\bf q}, {\bf f}\right\}}{\min} \; {\bf 1}_K^T 2^{\circ {\bf q}}\\ % \sum_{k = 1}^K \; 2^{q_k} \\ % E_k \left(p_k, {\bf V}, {\bf w}_k, {\bf D} \right)    \\
& \text{\it s.t.}
& & {\rm C_1}:  - \psi_k \left(\bar{\Gamma}_k + q_k \right) - \beta_k  +  \frac{f_k b_k}{B_W  f_k \bar{T}_k^{th} - \omega_k}  \leq 0, \; \forall k, \\
&&& {\rm C_2}: {\bf 1}^T_K {\bf f}  \leq F_T, \\
%&&& {\rm C_3}: 0 \leq 2^{q_k} \leq P_{k,max}, \; \forall k. \\
\end{aligned}
\end{equation}
where $\bar{\Gamma}_k \triangleq  \left(\log_2 (\alpha_{k,k}) - \log_2 \left(\bar{\eta}_k \left({\bf q}^{(t-1)} \right) +  \sum_{{\substack{j = 1, \;j \neq k}}}^K \alpha_{k,j} 2^{q_j^{(t-1)}} \right) \right)\leq \Gamma_k$. Therefore, any feasible solution for $\mathcal{P}_{7}$ is a feasible solution for $\mathcal{P}_{6}$. Accordingly, in the following we focus on $\mathcal{P}_{7}$ and resort to the KKT conditions to find the closed-form expressions for ${\bf p}^{(t)}$, i.e., $2^{\circ{\bf q}^{(t)}}$ and ${\bf f}^{(t)}$. Subsequently, the Lagrangian associated with $\mathcal{P}_{7}$ is given by

 % = 2^{\circ{\bf q}^{(t)}}
\begin{equation}
    \Upsilon \left({q}_k, {f_k}, {\vartheta_k}, {\mu} \right) = {\bf 1}_K^T 2^{\circ {\bf q}} + \sum_{k=1}^K \vartheta_k \left[ - \psi_k \left(\bar{\Gamma}_k + q_k \right) - \beta_k  +  \frac{f_k b_k}{B_W  f_k \bar{T}_k^{th} - \omega_k} \right] + \mu \left({\bf 1}^T_K {\bf f}  - F_T \right),
\end{equation}
%\sum_{k=1}^K \theta_k \left(2^{q_k} - P_{k,max} \right) - \sum_{k=1}^K \tau_k f_k
where the variables $\vartheta_k$ and $\mu$ are the non-negative Lagrange multipliers. Accordingly, the KKT conditions are given by

\vspace{-4mm}

\begin{IEEEeqnarray}{rCl} 
\IEEEyesnumber
\pdv{\Upsilon}{q_k} = (\log 2) 2^{q_k} - \vartheta_k \psi_k = 0, \; \forall k, \\
\pdv{\Upsilon}{f_k} = -\frac{\vartheta_k b_k \omega_k}{\left(B_W f_k \bar{T}^{th}_k - B_W \omega_k \right)^2} + \mu  = 0, \; \forall k, \\
\vartheta_k \left[ -\psi_k \left(\bar{\Gamma}_k + q_k \right) - \beta_k  +  \frac{f_k b_k}{B_W  f_k \bar{T}_k^{th} - \omega_k} \right] = 0, \; \vartheta_k \geq 0, \forall k,\\
\mu \left({\bf 1}^T_K {\bf f}  - F_T \right) = 0, \; \mu \geq 0.
\end{IEEEeqnarray}
Since $\mathcal{P}_7$ satisfies (28), all the $K$ IoTDs are served, i.e., $q_k > 0$ and $f_k > 0, \; \forall k$. Accordingly, the conditions (43), (44) and (46) imply $\vartheta_k > 0,\; \forall k$ and $\mu > 0$, which means that the computational capability at the BBU server is fully utilized, i.e., 

\begin{equation}
    {\bf 1}^T_K {\bf f}  = F_T.
\end{equation}
From another perspective, we can also obtain that ${\bf 1}_K^T{\bf f} < F_T$ would be sub-optimal, since at least the value of one $p_k$ can be further reduced by increasing the value of the corresponding $f_k$. Furthermore, $\vartheta_k > 0, \; \forall k$ implies that the latency constraint is always active, i.e.,

\begin{equation}
    -\psi_k \left(\bar{\Gamma}_k + q_k \right) - \beta_k  +  \frac{f_k b_k}{B_W  f_k \bar{T}_k^{th} - \omega_k}  = 0.
\end{equation}
This equation establishes a one-to-one relationship between the transmit power $p_k = 2^{q_k}$ and the number of cycles per second $f_k$ at the BBU server assigned to the $k$-th IoTD. Consequently, from (43) and (44), we obtain the expression for the optimal computational resource $f_k$ as

\begin{equation}
    f_k = \frac{1}{B_W \bar{T}^{th}_k} \left[\sqrt{\frac{(\log 2) b_k \omega_k 2^{q_k}}{\mu \psi_k}} + B_W \omega_k \right].
\end{equation}
By substituting (49) into (47) to obtain $\mu$ and by replacing $2^{q_k}$ with $p_k$, $f_k$ is further transformed into

\begin{equation}
    f_k = \frac{1}{B_W \bar{T}^{th}_k} \left[\frac{F_T - \sum_{k=1}^K \frac{\omega_k}{\bar{T}^{th}_k}}{\sum_{k=1}^K \frac{1}{B_W \bar{T}^{th}_k} \sqrt{\frac{(\log 2) b_k \omega_k p_k}{a_k}}} \sqrt{\frac{(\log 2) b_k \omega_k p_k}{a_k}} + B_W \omega_k \right].
\end{equation}
Finally, from (48), the optimal transmit power for the $k$-th IoTD is given by

\begin{equation}
    p_k = 2^{\left[\frac{1}{\psi_k} \left(\frac{f_k b_k}{B_W \bar{T}^{th}_k f_k - B_W \omega_k} - \beta_k \right) - \bar{\Gamma}_k\right]}.
\end{equation}
\noindent
\textit{\textbf{Lemma 3:}} Under the assumption that $\exists \; {\bf p}$ such that (28) is satisfied,  ${p_k}, \forall k$ obtained by (51) will converge to an optimal solution to $\mathcal{P}_6$ for a given $f_k, \; \forall k$.

\textit{Proof:} Refer to Appendix.

\noindent
\textit{\textbf{Lemma 4:}} Under the assumption that $\exists \; {\bf p}$ such that (28) is satisfied,  ${f_k}, \forall k$ given by (50) converges to a KKT point of $\mathcal{P}_6$.

\textit{Proof:} According to \textbf{\textit{Lemma 3}}, when $\exists \; {\bf p}$ such that (28) is satisfied, ${\bf p}$ obtained by (51) converges, i.e, ${\bf p}^{(t)} = {\bf p}^{(t-1)}$. Accordingly, upon convergence equality holds for $\bar{\Gamma}_k \leq {\Gamma}_k, \; \forall k$, which results in the equivalent KKT conditions for $\mathcal{P}_6$ and $\mathcal{P}_7$. Hence, $f_k, \; \forall k$ given by (50) converge to KKT point of $\mathcal{P}_6$. $\blacksquare$ 

\begin{algorithm}[ht]
 \begin{algorithmic}[1]
 %\Procedure{Euclid}{$a,b$}\Comment{The g.c.d. of a and b}
\STATE \textbf{Input}: ${\bf p}^{(0)}$, $\varpi^{(0)}$.
\STATE Initialize $t \gets 1$;
\REPEAT
%\STATE $z_k^{(t)} =  \frac{p_k^{(t-1)} \alpha_{k,k}}{\sum_{j \neq k}^K p_j^{(t-1)} \alpha_{k,j} + \bar{\eta}_k}, \; \forall k$;
\STATE Update $\psi_k^{(t-1)}$, $\beta_k^{(t-1)}$ using (35),  $\forall k$;
\STATE Update $f_k^{(t)}$ using (50),  $\forall k$;
\STATE Update $p_k^{(t)}$ using (51),  $\forall k$; 
\STATE $t \gets t + 1$.
\UNTIL{convergence}
\STATE \textbf{Output}: $p_k$, $f_k, \;\forall k$.
%\EndProcedure
\end{algorithmic}
\caption{Iterative algorithm to solve $\mathcal{P}_6$}
\label{AG1}
\end{algorithm}

For clarity, we summarize the above procedure in Algorithm~\ref{AG1}, which describes  the  framework  to  obtain  the transmit power and computational resource for the $K$ IoTDs. Since $\mathcal{P}_6$ satisfies the conditions in (33), its solution will converge to the KKT point of $\mathcal{P}_4$, which accordingly gives a local minimum of $\mathcal{P}_{4}$  \cite[Corollary 1]{marks1978general}. Hence, according to \textbf{\textit{Lemma 3}} and \textbf{\textit{Lemma 4}}, Algorithm~\ref{AG1} converges to a local minimum of $\mathcal{P}_{4}$.

\subsubsection{Solution for the problem $\mathcal{P}_5$}

$\mathcal{P}_5$ is a non-convex problem due to its constraints ${\rm C_1}$ and ${\rm C_3}$. Noting that there is only a single integer variable to be optimized, we resort to the line-search method to find the optimal $\varpi$ over the feasible set. Accordingly, $\varpi$ is given by

\begin{equation}
    \varpi = \underset{\tilde{\varpi}}{\argmax}  \; \left\{\tilde{\varpi} \in \mathbb{Z}_{>0} : \xi_k^{TL} \left({\bf p}, \tilde{\varpi} \right)  +  \frac{2 b_k  R \varpi}{C_F \log_2(M)} \leq  \tilde{\mathcal{T}}_k^{th}, \; \forall k, \tilde{\varpi} \leq \frac{C_{F}}{2 B_W R} \right\}.
\end{equation}

\subsubsection{Overall algorithm for the problem $\mathcal{P}_2$}

Algorithm~\ref{AG2} summarizes the overall algorithm to solve $\mathcal{P}_2$. Specifically, for a given feasible ${\bf p}$, ${\bf f}$ and $\varpi$, the algorithm starts by obtaining ${\bf W}$ using (23). Subsequently, for the obtained ${\bf W}$ and a fixed $\varpi$, ${\bf p}$ and ${\bf f}$ are updated using Algorithm 1. Finally, for the obtained ${\bf W}, {\bf p}$, and ${\bf f}$, we find a feasible $\varpi$ using (52) for the next iteration. As the objective of $\mathcal{P}_2$ is decreasing in each iteration owing to Algorithm~\ref{AG1}, Algorithm~\ref{AG2} converges to a local minimum.

\begin{algorithm}[ht]
 \begin{algorithmic}[1]
 %\Procedure{Euclid}{$a,b$}\Comment{The g.c.d. of a and b}
\STATE \textbf{Input}: ${\bf p}^{(0)}$, $\varpi^{(0)}$. %, predefined stopping criteria \{$ITR_{max}$\}.
\STATE Initialize $t \gets 1$;
\REPEAT
\STATE Update ${\bf W}^{(t)}$ using (23);
\STATE Update ${\bf p}^{(t)}$ and ${\bf f}^{(t)}$ using Algorithm 1;
\STATE Update $\varpi^{(t)}$ using (52); 
\STATE $t \gets t + 1$. %using Algorithm 2;
\UNTIL{convergence}
\STATE \textbf{Output}: ${\bf W}$, ${\bf p}$, ${\bf f}$, ${\varpi}$.
%\EndProcedure
\end{algorithmic}
\caption{Overall algorithm to solve $\mathcal{P}_2$}
\label{AG2}
\end{algorithm}

\subsection{Low-Complexity Implementation for the Joint Optimization based on Deep Learning}

\begin{figure}[htb]
       \centering
           \includegraphics[scale=0.375]{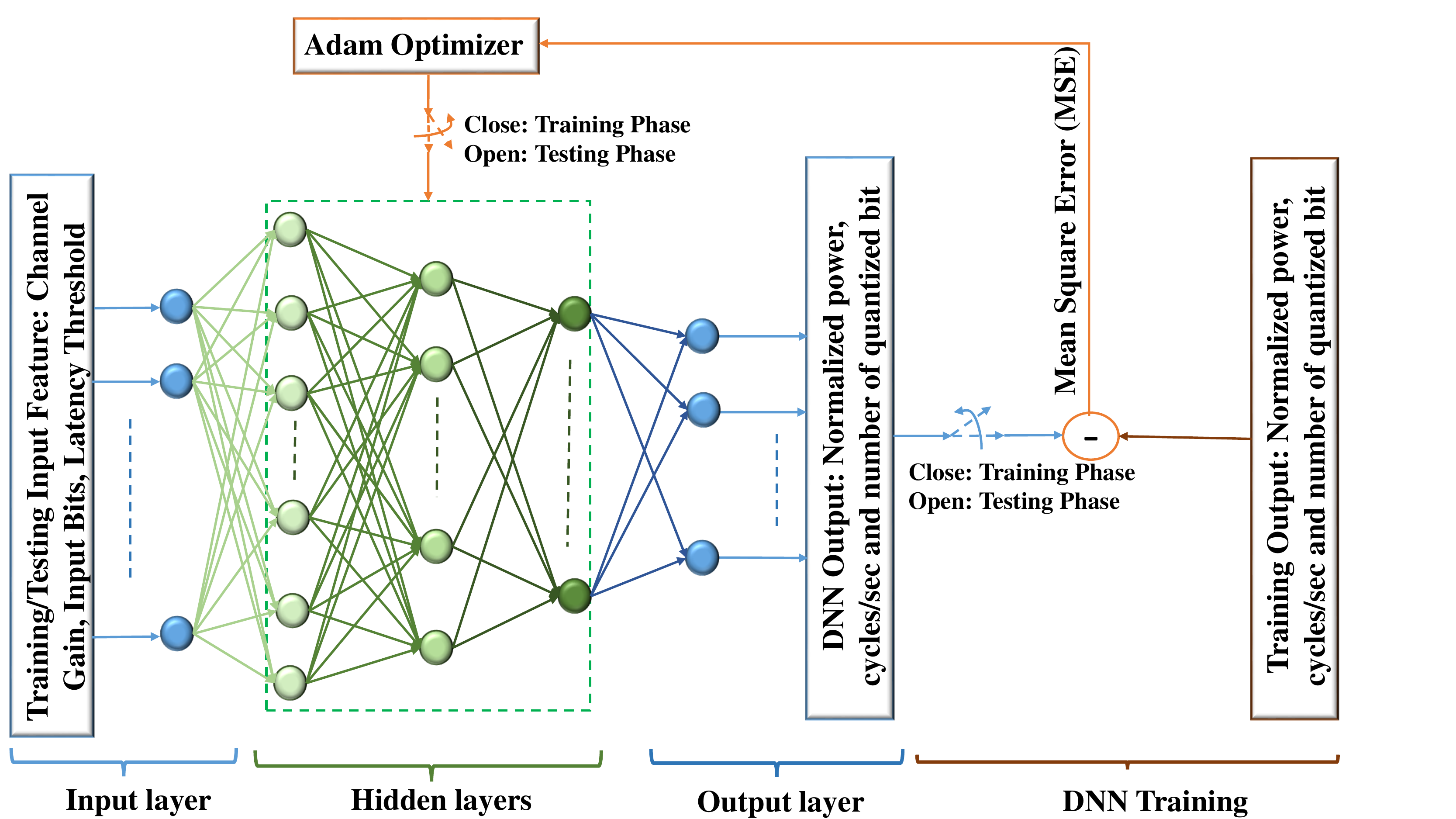}
\caption{\small DNN architecture for the proposed supervised deep learning with the training and testing phase.}
    \label{DNNA}
\end{figure}

Although Algorithm~\ref{AG2} obtains near-optimal solutions for $\mathcal{P}_2$, it involves an interleaved loop structure which can limit its practicability in terms of the real-time processing. Accordingly, in  this  section, we present a supervised deep learning method using the DNN to approximate the proposed Algorithm 2, such that by passing the input operating parameters of Algorithm 2 through a trained DNN gives a feasible output for the resource allocation for the C-RAN network with much reduced execution time. Furthermore, training the DNN is fairly convenient as the training samples can easily be obtained by running Algorithm~\ref{AG2} offline \cite{SunDeep2018}.
 Next, we describe the DNN architecture used in our work, as shown in Fig.~\ref{DNNA}. Specifically,  the DNN consists of a)  one input layer of $3K$  neurons by aligning the computation bits, the latency thresholds and the effective channel at BBU for $K$ IoTDs into a column vector defined as $ {\bf x}_D \triangleq \left[{\bf b}^T, {\mathcal{\bf T}}_{th}^T, {\bf 1}^T_K \left({\bf V}_D {\bf H} \odot {\bf I}\right)\right]^T$, where ${\bf b} \triangleq \left[b_1, \dots, b_k \right]^T$ and ${\mathcal{\bf T}}_{th} \triangleq \left[{\mathcal{T}}_1^{th}, \dots, {\mathcal{T}}_K^{th} \right]^T$, b) one output layer of $2K + 1$ neurons corresponding to the transmit powers and computational resources for the $K$ IoTDs, and the quantization bit allocation, jointly defined by the column vector ${\bf y}_{D} \triangleq \left[{\bf p}^T, {\bf f}^T, \varpi \right]^T$, and c) $L - 1$ fully connected hidden layers. Let $\mathcal{L} \triangleq \left\{0,  \dots, L \right\}$ represent the set of layers, where $l = 0$ and $l = L$ denote the input and output layers, respectively. The number of neurons in each layer $l \in \mathcal{L}$ is denoted by $n_l$, and accordingly, we have $n_0 = 3K$ and $n_L = 2K + 1$. For each hidden layer $l$, the output ${\bf y}_l \in \mathbb{R}^{n_l \times 1}$ is calculated as

\begin{equation}
    {\bf y}_l =  {\rm ReLU} \left({\bf Q}_l {\bf y}_{l - 1} + b_l \right), \; l \in \left\{1, \dots, L-1 \right\},
\end{equation}
where ${\bf y}_{l - 1} \in \mathbb{R}^{n_{l-1} \times 1}$ is the output of the $(l - 1)$-th layer with ${\bf y}_0 = {\bf i}_D$, ${\bf Q}_l \in \mathbb{R}^{n_l \times n_{l-1}}$ and ${\bf b}_l \in \mathbb{R}^{n_l \times 1}$ are respectively the weight matrix and bias vector at the $l$-th layer, and ${\rm ReLU} \left(x \right) = \max \left(x,0 \right)$ is the Rectified Linear Unit function, which introduces nonlinearity to the network. Accordingly, the deep learning method involves 

\begin{enumerate}
    \item Obtaining the training data (${\rm Train_D}$), i.e., the training input ${\bf x}_D$ and the training output ${\bf y}_D$ from Algorithm~\ref{AG2}.
    
    \item Normalizing ${\rm Train_D}$ such that ${\rm Train_D} \in \left[0,1\right]$.
    
    \item Deploying the mini-batch gradient descent based on Adam optimizer to train the DNN \cite{learningSurvey2019, DeepSVKim2018, chollet2015keras, kingma2014adam, DeepSVKim2018} as shown in Fig.~\ref{DNNA} (Training phase), which effectively minimizes the mean square error (MSE) given by 
    
     \begin{equation}
         {\rm MSE} = \frac{\sum_{b=1}^{B_M} \sum_{i=1}^{2K+1} \left( o_{L,i,b}- o_{D,i,b}\right)^2}{B_M\left(2K + 1\right)},
     \end{equation}
     where $B_M$ is the number of mini-batches, $o_{L,i,b}$ and  $o_{D,i,b}$ are the  outputs at the $i$-th neuron of the $L$-th layer and the corresponding training output, respectively, for the $b$-th mini-batch.
    
    \item After the training phase, the DNN is used to obtain the desired output based on the test data, which can be real-time data from the C-RAN network, as shown in Fig.~\ref{DNNA} (Testing phase).
  
\end{enumerate}

\section{Numerical Results}

In  this  section,  we  evaluate  the  performance of our proposed approach via Monte-Carlo simulations. Unless otherwise stated, we consider a network composed of $N = 128$ antennas randomly deployed on a wall in a $10 \; {\rm m} \times 10 \; {\rm m} \times 10 \; {\rm m}$ indoor room as shown in Fig.~\ref{SetUP}. Furthermore, there are $K = 10$ single-antenna IoTDs uniformly distributed inside the room. The number of bits $b_k$ and latency threshold $\mathcal{T}_k^{th}$ for each IoTD's computational task $C_k$ are  randomly assigned  between $10 \; {\rm kbs}$ to $20 \; {\rm kbs}$ and $0.5 \; {\rm s}$ to $1 \; {\rm s}$, respectively.  The computation bits are encoded using the QPSK modulation, i.e., $M = 4$. For the sake of simplicity, the number of CPU cycles needed for completing $C_k$ is set as a linear function of $b_k$, i.e.,  $\omega_k = \eta b_k$, with $\eta = 50$ \cite{Barbarossa2013}.  The carrier frequency of the wireless links is taken to be $f_c = \frac{c}{\lambda} = 1.5 \; {\rm GHz}$ with a transmission bandwidth of $B_W = 180 \; {\rm KHz}$, where $c = 3 \times 10^8 \; {\rm m/s}$. Furthermore, the channel parameters are given as $\xi = 3.7$, $\kappa = \left(13 - 0.03 \; d_{n,k}[m]\right) {\rm dB}$ and $\tau_{n,k} = 6\; {\rm dB}, \; \forall n,k$ \cite{tse2005fundamentals, DeepUSVXu2019, LISWaad2018}. The transmit power constraint for each user is $P_{k, max} = 0 \; {\rm dBm}$. The power spectral density of the background noise at the xL-MIMO RRH is assumed to be $-169 \; {\rm dBM/Hz}$, and the noise figure due to the receiver processing is $7 \; {\rm dB}$ \cite{zhang2015CRAN}. Lastly, it is assumed that  the BBU server has a computational capability of $F_T =  15 \; {\rm MHz} \; {\rm cycles/s}$ with a fronthaul capacity of $C_{F} = 100 \; {\rm MHz}$. The above choice of parameters guarantees the non-emptiness of the feasible set for $\mathcal{P}_3$, where ${p}_k^{(0)}, \; \forall k$ and $\varpi^{(0)}$ are selected randomly within the feasible sets $\Psi$ and $\mathcal{D}$, respectively.

 \begin{figure}[!t]
       \centering
               \includegraphics[scale=0.225]{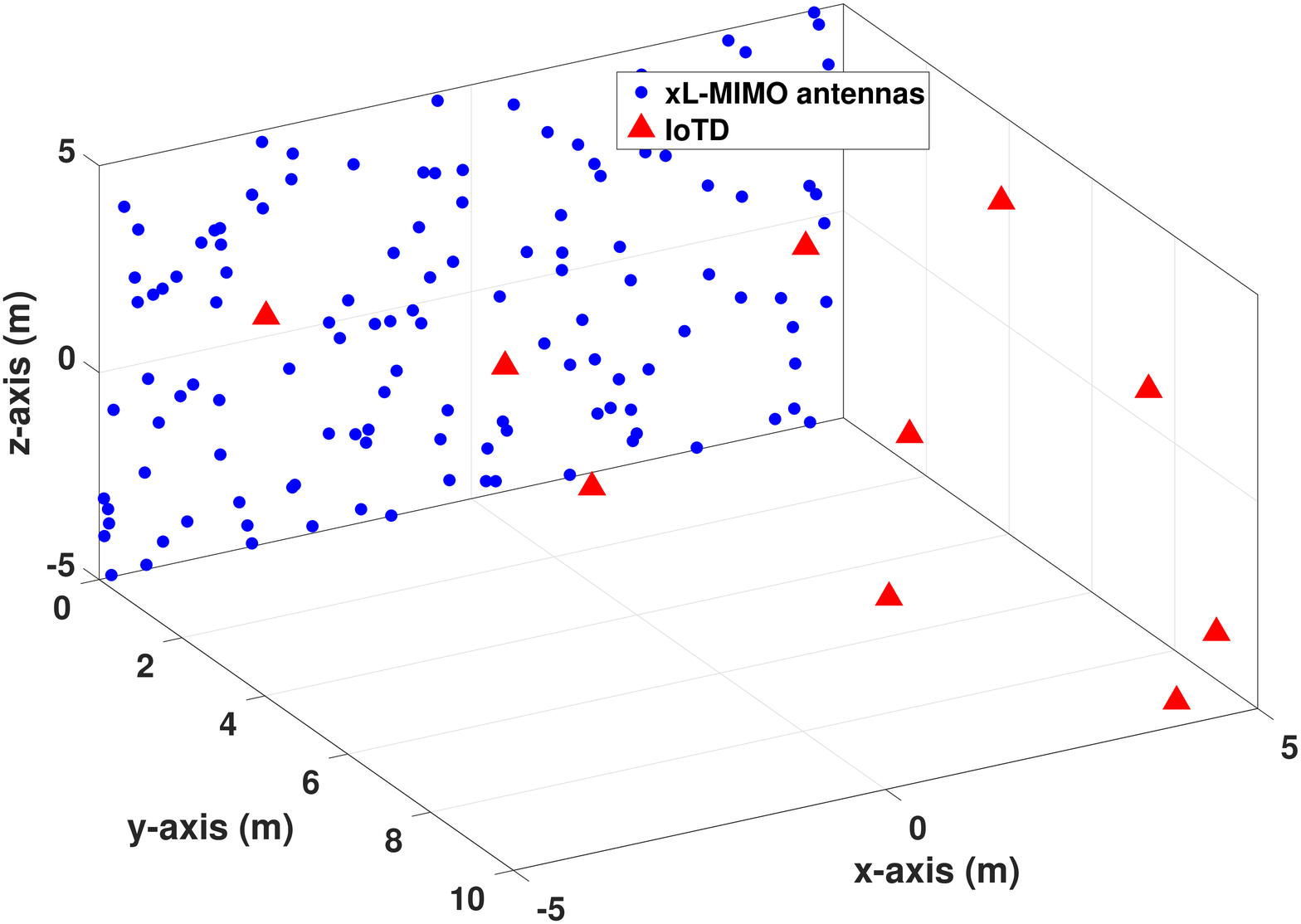}
\caption{\small Simulation set-up with $N$ antennas (blue circles) deployed on a wall and $K$ IoTDs (red triangles) distributed in an indoor room.}
    \label{SetUP}
\end{figure}
\subsection{Performance per IoTDs' Distribution}

To gain insights from the communication and computational resource allocations by the proposed algorithm, we firstly consider the resource allocation for a particular distribution of the IoTDs and channel realization. In Fig.~\ref{OVA}, we illustrate the obtained communication (transmit power $p_k, \; \forall k$, first sub-figure) and computational (normalized number of CPU cycles $\frac{f_k}{F_T}, \; \forall k$, second sub-figure) resources assigned to each IoTD with respect to the corresponding effective channel gains at the BBU, i.e., $\left|{\bf h}^e_k \right|  = \left| {\bf 1}^T_K \left({\bf V}_D {\bf H} \odot {\bf I}\right) \right|, \; \forall k$ (third sub-figure), the number of computation bits ($b_k, \; \forall k$, fourth sub-figure) and the latency thresholds ($\mathcal{T}_k^{th},\; \forall k$, fifth sub-figure). In the fifth sub-figure, we also plot the overall latency $ \xi_{k},\; \forall k$ computed using (16).

\begin{figure}[!t]
       \centering
              \includegraphics[scale=0.30]{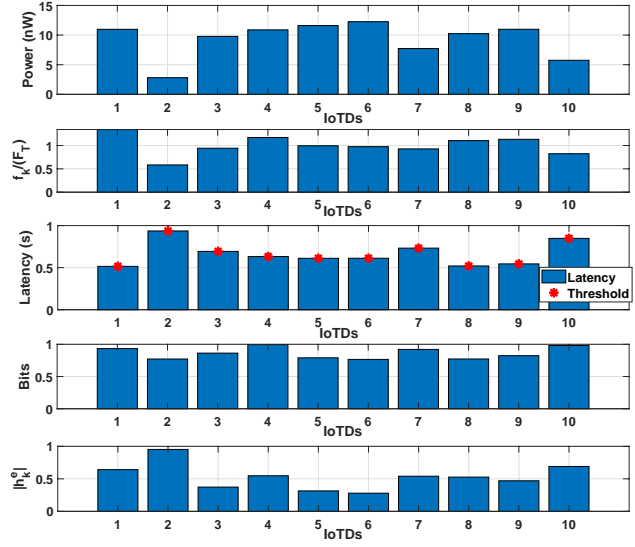}
\caption{\small Optimal transmitted power $p_k$, normalized CPU cycles $f_k/F_T$ and overall latency $\xi_{k}$, with respect to the effective channel gain ${\bf h}^e_k$, the number of transmit bits $b_k$ and the latency threshold $\mathcal{T}^{th}_k$ corresponding to each IoTD.}
    \label{OVA}
\end{figure}

As observed, the proposed algorithm assigns a higher transmit power and CPU cycles to IoTDs  with a poor effective channel gain (IoTD 5,6), a larger number of computation bits (IoTD 1,4) or a stringent latency constraint (IoTD 8,9). An interesting observation is that, with similar channel gains, the latency constraint dominates over the number of computation bits in determining the allocation of the communication and computational resources as observed for IoTD 7 and 8. This demonstrates that the latency constraints play a crucial role in the computation offloading for the IoTDs. Furthermore, it is seen that the computational tasks of all the IoTDs are executed within the respective latency constraint. 

\vspace{-4mm}
\subsection{Joint versus Disjoint Optimization}

\begin{figure}[!htb]
 \centering
   \subfigure[$P_{\rm sum}$ vs $N$]
    {
        \includegraphics[scale=0.25]{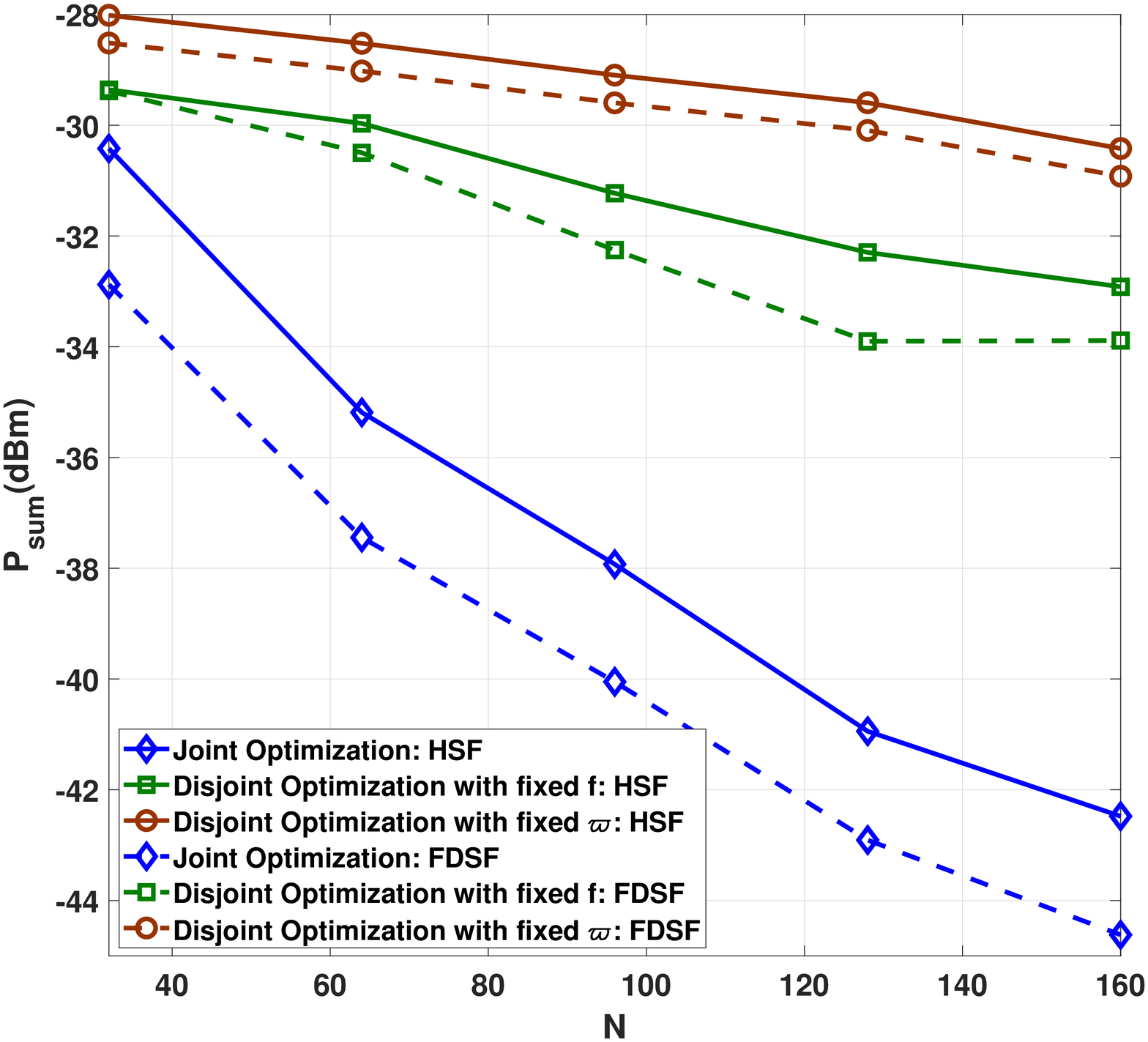} 
        } \hskip -1.95ex
    \subfigure[$P_{\rm sum}$ vs $\eta$] 
    {
       \includegraphics[scale=0.25]{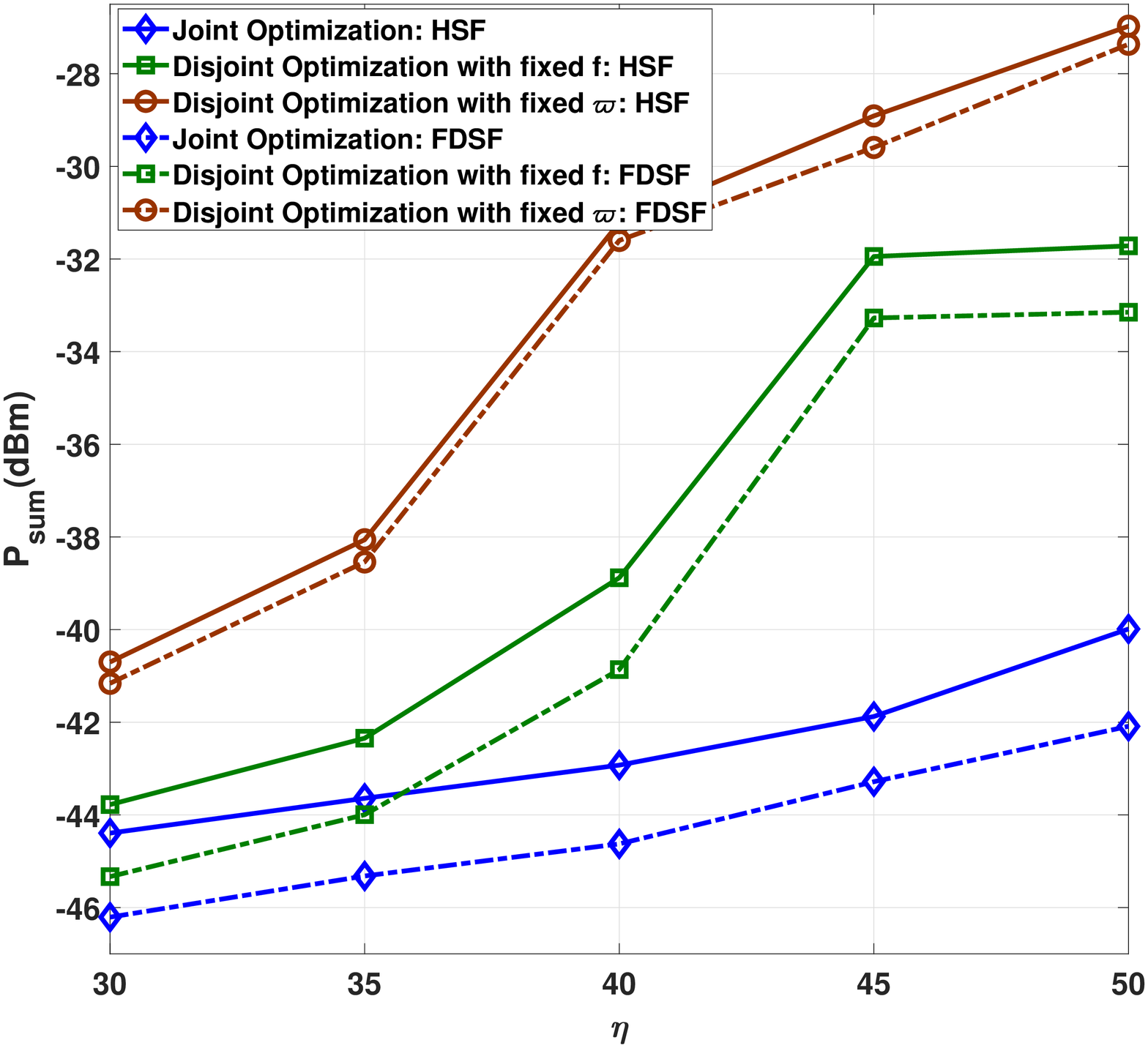} %{figG.eps}
 } 
    \caption{Total transmit power $P_{\rm sum}$ versus the number of antennas $N$ and  the computational load $\eta$ for the proposed joint optimization and the disjoint optimizations.}
    \label{PSUM}
\end{figure}

In this section, we evaluate the merit of the proposed algorithm with two benchmark algorithms: 1) Disjoint optimization with fixed ${\bf f}$: Solving $\mathcal{P}_2$ with Algorithm 2 where $p_k, \; \forall k$ and $\varpi$ are optimized with $f_k = \frac{\omega_k F_T}{\sum_{k=1}^K \omega_k}, \; \forall k$, which meets the computational rate constraint $F_T$ with equality \cite{Barbarossa2015}, and 2) Disjoint optimization with fixed ${\varpi}$: Solving $\mathcal{P}_2$ with Algorithm 2 where $p_k, \; \forall k$ and $f_k, \; \forall k$ are optimized with the number of quantization bits $\varpi$  fixed at $\varpi = \left\lceil{\frac{C_F}{4 B_W L}}\right\rceil$, i.e., half of the maximum feasible $\varpi$. We assess the usefulness of the algorithms with respect to the number of antennas $N$ at the xL-MIMO RRH and the computational load given by the ratio $\eta = \frac{\omega_k}{b_k}$ between the required number of CPU cycles $\omega_k$ and number of computation bits $b_k$ \cite{Barbarossa2015}. 

Fig.~\ref{PSUM}(a) shows the total transmit power of the IoTDs with respect to $N$ for $\eta = 50$, obtained using Algorithm 2 and the disjoint optimization algorithms, with both the HSF and the FDSF. It can be observed that the proposed joint optimization algorithm yields a considerable gain compared to the disjoint optimization algorithms, where deploying a large number of antennas results in a decrease in the total transmit power. This decrease in the total trasmit power is because of the array gain, which is proportional to $N$, resulting in a decrease in the required transmit power of each IoTD \cite{ngo2013energy}. Furthermore, this explains the use of the xL-MIMO with a large $N$ to minimize the power drainage of IoTDs and consequently, extend their battery life. 

Next, Fig.~\ref{PSUM}(b) presents the total transmit power of the IoTDs with respect to $\eta$  for $N = 128$  and $\omega_k = \eta b_k, \; \forall k$, obtained using the algorithms, with both the HSF and the FDSF. Specifically, $\eta$ is varied with $b_k$ and $\mathcal{T}^{th}_k$ randomly set between $10 \; {\rm kbs}$ to $20 \; {\rm kbs}$ and  $0.5 \; {\rm s}$ to $1 \; {\rm s}$, respectively. It can be observed that the proposed joint optimization algorithm outperforms the disjoint optimization algorithms for the computational tasks with a stringent computational requirement. Finally, it can be seen from Fig.~\ref{PSUM} that there is a performance loss for the HSF compared to the FDSF owing to a loss in the spectral efficiency for the hybrid architecture \cite{jointKim2019, yu2016alternating, sohrabi2016hybrid}.

\subsection{Deep Neural Network Evaluation}

\subsubsection{Impact of small-scale fading on DNN training}

\begin{figure}[!htb]
 \centering
   \subfigure[pdf for $p_k$ ]
    {
        \includegraphics[scale=0.25]{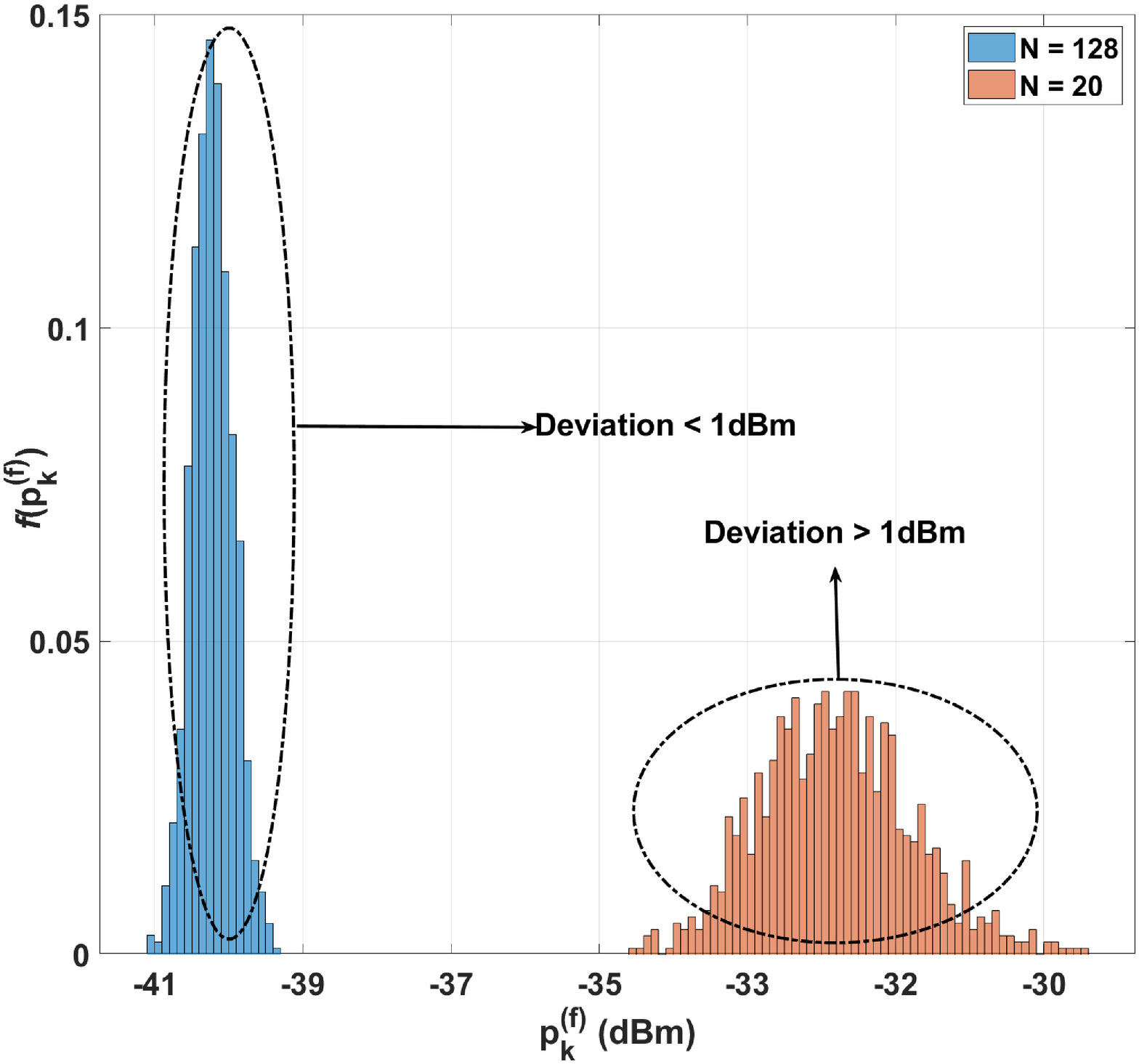} %{figH.eps}
    } \hskip -4.0ex
    \subfigure[pdf for $\frac{f_k}{F_T}$] 
    {
       \includegraphics[scale=0.25]{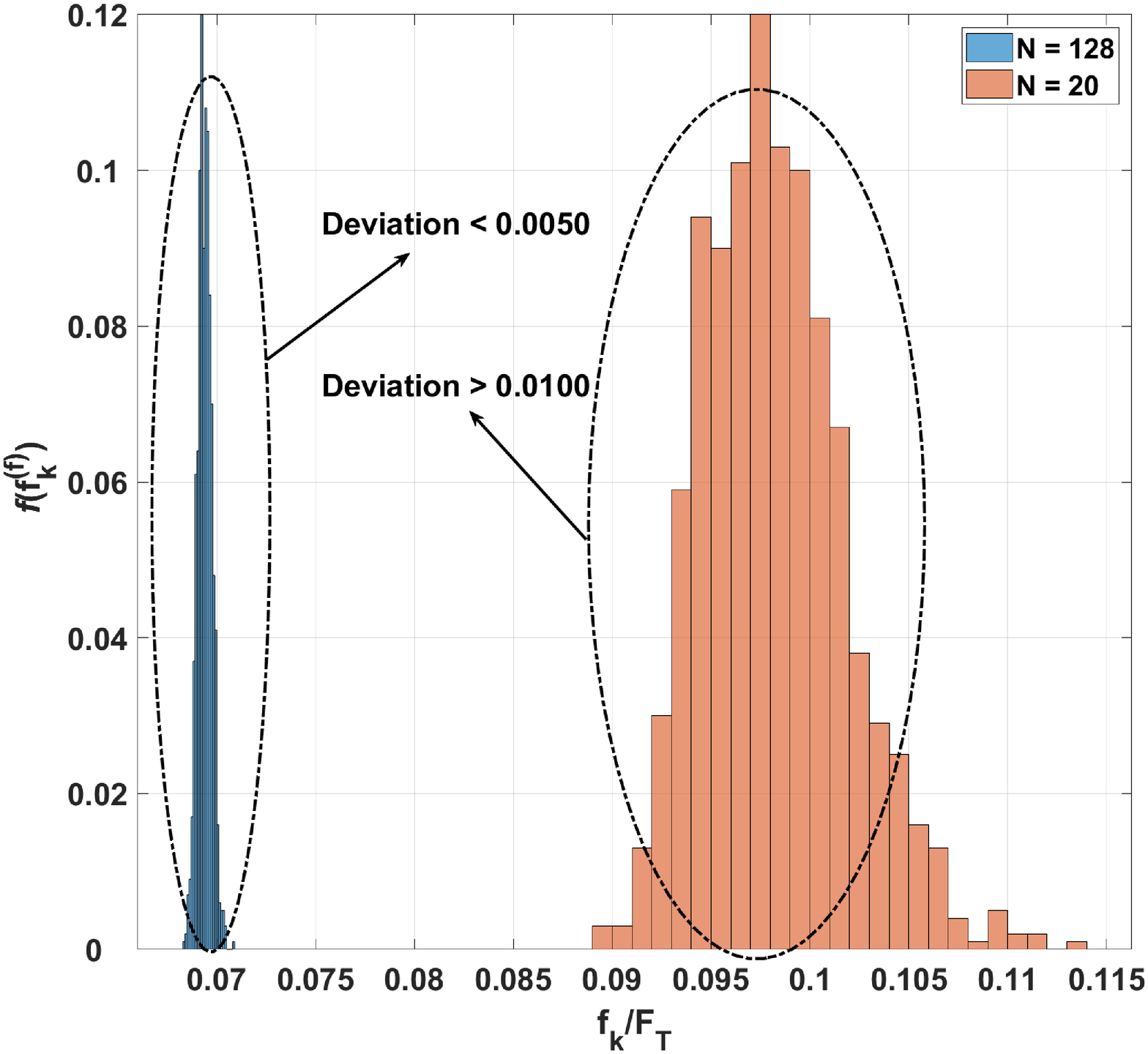} %{figG.eps}
 } 
    \caption{\small Probability density function of the optimal transmit power $p_k$ and the normalized CPU cycles $\frac{f_k}{F_T}$ assigned to an IoTD.}
    \label{PDF}
\end{figure}

We begin by evaluating the impact of the small-scale fading on the resource allocation for the computation offloading. Accordingly, for a fixed number of computation bits $(b_k, \; \forall k)$ and latency thresholds $(\mathcal{T}_k^{th}, \; \forall k)$, we ran the proposed algorithm for $10^3$ channel initializations for a fixed distribution of the IoTDs. In Fig.~\ref{PDF}(a) and \ref{PDF}(b), we show the pdf for the optimal transmit power  $p_k$ and normalized number of CPU cycles $\frac{f_k}{F_T}$ assigned to an IoTD across $10^3$ channel realizations. It can be seen that for $N = 128$, $p_k$\footnote{As analyzed in \cite{ngo2013energy}, a larger number of antennas at the xL-MIMO RRH is seen to decrease the required transmit power of the IoTD, thereby further minimizing the power consumption and the total transmit power as seen in Fig.5(a).} and $\frac{f_k}{F_T}$ have a significantly lesser deviation compared to that for $N = 20$. Hence, for a fixed $b_k$ and $\mathcal{T}_k^{th}, \; \forall k$, these results demonstrate that the proposed HSF with a large number of antennas at the xL-MIMO RRH reduces the impact of the small-scale fading on the resource allocation for the IoTDs as explained in Section 3. Consequently, the BBU needs to update the operating parameters depending only on the large-scale fading of the IoTDs. Additionally, this explains the use of the effective channel at the BBU, representing the large-scale fading corresponding to each IoTD as shown in (19), as an input parameter to train the proposed DNN along with $b_k$ and $\mathcal{T}_k^{th}, \; \forall k$.  

\subsubsection{DNN training and testing}

We implemented the proposed DNN scheme with the Keras machine learning toolkit where Adam optimizer was used for minimizing the MSE during the training phase \cite{chollet2015keras, kingma2014adam, DeepSVKim2018}. Accordingly, we consider three hidden layers with $128$, $64$ and $32$ neurons for $l = 1, 2$ and $3$, respectively. We collected $50000$ $Train_{D}$ sets, which are split in the ratio of $9:1$ for the training and testing of the DNN. Fig.~\ref{loss} shows the training and testing losses with respect to the number of epochs, which can be seen to converge within $20$ epoch. 

\begin{figure}[!t]
       \centering
       \includegraphics[scale=0.20]{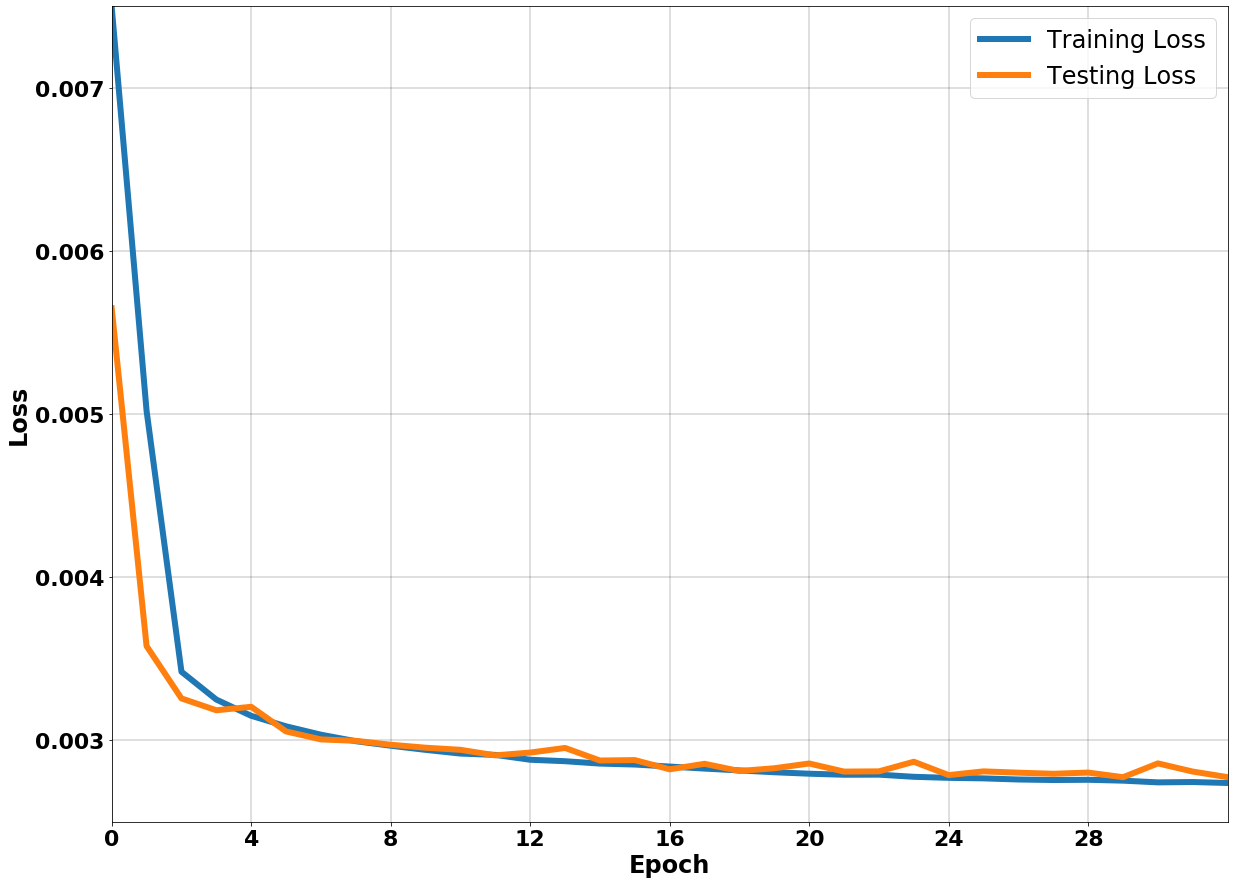}
\caption{\small Training and testing losses versus epoch  for the DNN based learning.}
    \label{loss}
\end{figure}

\begin{figure}[!t]
       \centering               
       \includegraphics[scale=0.28]{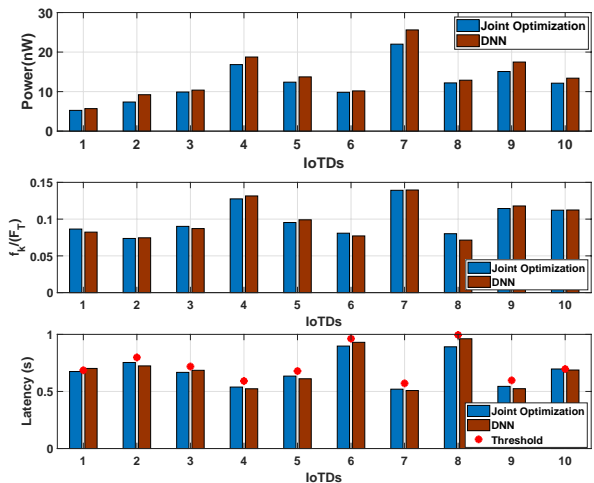}
\caption{\small Transmitted power $p_k$, normalized CPU cycles $f_k/F_T$ and overall latency $\xi_{k}$ obtained from the proposed algorithm (elapsed time: $118 \; {\rm ms}$) and the DNN (elapsed time: $1 \; {\rm ms}$) with respect to the effective channel gain ${\bf h}^e_k$, the number of transmit bits $b_k$ and the latency threshold $\mathcal{T}_k^{th}$ corresponding to each IoTD.}
    \label{DNN}
\end{figure}

%, denoted by ${\rm Opt}$ and ${\rm DNN}$, respectively
Next, in Fig.~\ref{DNN}, we show the communication (transmit power $p_k, \; \forall k$, first sub-figure) and computational (normalized number of CPU cycles $\frac{f_k}{F_T}, \; \forall k$, second sub-figure) resources, and the overall latency along with the respective latency thresholds ($\xi_{k}$ and $\mathcal{T}_k^{th},\; \forall k$, third sub-figure) obtained from the proposed joint optimization through Algorithm 2 and the trained DNN for an IoTDs' distribution and channel realization. As observed, although the transmit power of the IoTDs obtained from the DNN is marginally higher than that obtained from Algorithm 2, the DNN is able to emulate the performance of Algorithm 2 in allocating the resources to the IoTDs, while satisfying the latency requirement. Finally, in Fig.~\ref{cdf}, we evaluate the total transmit power performance of the DNN based approach in the testing phase compared to the proposed joint optimization and the disjoint optimizations described in the previous sub-section. The cumulative distribution function (CDF) for $P_{sum}$ in Fig.~\ref{cdf} is obtained over 5000 testing data sets \cite{SunDeep2018}. It is observed that the total transmit power of the IoTDs obtained from the trained DNN is very close to that obtained from the proposed joint optimization while significantly outperforming the disjoint optimizations. Furthermore, we measured the elapsed time for the computation of the optimal resource allocations through the proposed joint optimization and the trained DNN, where Intel core i7-6700 CPU@$3.40 \; {\rm GHz}$ and $16.00 \; {\rm GB}$ RAM are used. The average elapsed time per computation corresponding to the proposed joint optimization and the trained DNN was found to be $118 \; {\rm ms}$ and $1 \;{\rm ms}$, respectively, which highlights the practicability of the proposed deep learning method. 

\begin{figure}[!t]
       \centering
       
               \includegraphics[scale=0.245]{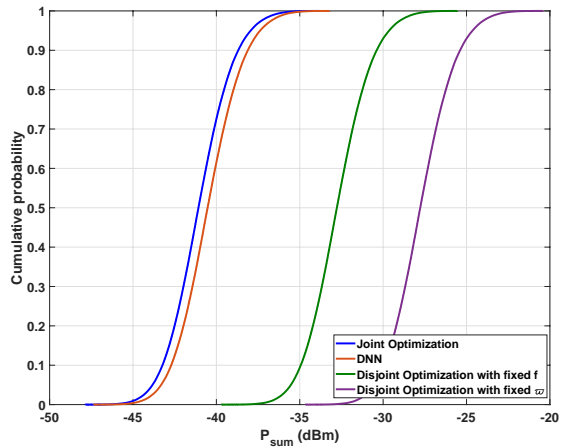}
\vspace{-2mm}
\caption{\small CDF for $P_{sum}$ achieved by the proposed joint optimization, the DNN and the disjoint optimizations.}
    \label{cdf}
\end{figure}

\section{Conclusion}

In this paper, we have formulated a computation offloading problem for IoT applications with a latency constraint in an uplink xL-MIMO C-RAN. The constructed optimization problem that minimizes the total transmit power of the IoTDs while satisfying the latency requirement is found to be non-convex. With the HSF matrix obtained locally at the xL-MIMO RRH, the joint optimization on the baseband combiner, the communication and computational resource allocations, and the number of quantization bits at the BBU is solved with the alternating optimization based on the concepts of the MMSE metric, the successive inner convexification and the linear-search method, respectively. Furthermore, a supervised  deep learning method using the DNN is deployed as an efficient solution. Numerical results validate the effectiveness of the proposed joint optimization scheme, which outperforms two benchmarks based on disjoint optimization. The efficiency of the DNN-based method is also verified.

  \appendix

  \section{Proof for Lemma 3}
  \label{FirstAppendix}
  
 For a given ${f}_k, \; \forall k$, $\mathcal{P}_6$ can be further transformed into, 

\begin{equation}
\begin{aligned}
& \mathcal{P}_{8}: && \underset{\left\{{\bf q} \preccurlyeq \log_2\left({\bf P}_{max}\right)\right\}}{\min} \; {\bf 1}_K^T 2^{\circ {\bf q}}\\ % E_k \left(p_k, {\bf V}, {\bf w}_k, {\bf D} \right)    \\
& \text{\it s.t.}
& & {\rm C_1}:  - \psi_k \left({\Gamma}_k + q_k \right) - \beta_k  +  \frac{f_k b_k}{B_W  f_k \bar{T}_k^{th} - \omega_k}  \leq 0, \; \forall k. \\
%&&& {\rm C_3}: 2^{q_k} \leq P_{k,max}, \; \forall k, \\
\end{aligned}
\end{equation}
By substituting $p_k = 2^{q_k}$ and assuming $\exists \; {\bf p}$ such that (28) is satisfied, i.e., ${\bf p} \preccurlyeq {\bf P}_{max}$, $\mathcal{P}_{8}$ reduces to the following problem:

\begin{equation}
\begin{aligned}
& \mathcal{P}_{9}:  \underset{\left\{{\bf p} \succeq {\bf I}\left({\bf p}\right)\right\}}{\min} \;  {\bf 1}_K^T {\bf p},
%& \text{\it s.t}
%& & {\rm C_1}: {\bf p} \geq {\bf I}({\bf p}),
\end{aligned}
\end{equation}
where the function  ${\bf I}({\bf p}) \triangleq \left[{I}_1({\bf p}),  \dots, {I}_K({\bf p}) \right]^T \in \mathbb{R}^{K \times 1}$ is a \textit{standard interference function} \cite{Yates1995PC}, with each entry given by

\begin{equation}
    I_k({\bf p}) =  2^{\left[\frac{1}{\psi_k} \left(\frac{f_k b_k}{B_W \bar{T}^{th}_k f_k - B_W \omega_k} - \beta_k \right) - {\Gamma}_k\right]}.
\end{equation}
$\mathcal{P}_9$ is the well-known power control problem \cite{Yates1995PC}, which has an optimal solution obtained through the \textit{standard power control algorithm}, given by ${\bf p}^{(t)} = {\bf I}\left({\bf p}^{(t-1)}\right)$. According to \cite[{Corollary 4.1 {\rm and} 4.2}]{zhang2015CRAN}}, given a feasible $\mathcal{P}_8$, ${\bf p}^{(t)} = {\bf I}\left({\bf p}^{(t-1)}\right) \preccurlyeq {\bf p}^{(t-1)}$ will converge to the optimal solution to $\mathcal{P}_8$ with any initial point ${\bf p}^{(0)} \succeq 0$. Hence, (51) which is in the form of ${\bf p}^{(t)} = {\bf I}\left({\bf p}^{(t-1)}\right)$ converges to an optimal solution for $\mathcal{P}_6$ for a given $f_k, \; \forall k$. $\blacksquare$ 

\vspace{-2mm}

\bibliographystyle{IEEEtran}
\bibliography{main.bib}

\begin{thebibliography}{10}
\providecommand{\url}[1]{#1}
\csname url@samestyle\endcsname
\providecommand{\newblock}{\relax}
\providecommand{\bibinfo}[2]{#2}
\providecommand{\BIBentrySTDinterwordspacing}{\spaceskip=0pt\relax}
\providecommand{\BIBentryALTinterwordstretchfactor}{4}
\providecommand{\BIBentryALTinterwordspacing}{\spaceskip=\fontdimen2\font plus
\BIBentryALTinterwordstretchfactor\fontdimen3\font minus
  \fontdimen4\font\relax}
\providecommand{\BIBforeignlanguage}[2]{{%
\expandafter\ifx\csname l@#1\endcsname\relax
\typeout{** WARNING: IEEEtran.bst: No hyphenation pattern has been}%
\typeout{** loaded for the language `#1'. Using the pattern for}%
\typeout{** the default language instead.}%
\else
\language=\csname l@#1\endcsname
\fi
#2}}
\providecommand{\BIBdecl}{\relax}
\BIBdecl

\bibitem{popli2019survey}
S.~Popli, R.~K. Jha, and S.~Jain, ``{A Survey on Energy Efficient Narrowband
  Internet of Things (NBIoT): Architecture, Application and Challenges},''
  \emph{IEEE Access}, vol.~7, pp. 16\,739--16\,776, 2019.

\bibitem{feltrin2019narrowband}
L.~Feltrin, G.~Tsoukaneri, M.~Condoluci, C.~Buratti, T.~Mahmoodi, M.~Dohler,
  and R.~Verdone, ``{Narrowband IoT: A Survey on Downlink and Uplink
  Perspectives},'' \emph{IEEE Wireless Communications}, vol.~26, no.~1, pp.
  78--86, 2019.

\bibitem{Kumar2013}
\BIBentryALTinterwordspacing
K.~Kumar, J.~Liu, Y.-H. Lu, and B.~Bhargava, ``{A Survey of Computation
  Offloading for Mobile Systems},'' \emph{Mobile Networks and Applications},
  vol.~18, no.~1, pp. 129--140, Feb 2013. [Online]. Available:
  \url{https://doi.org/10.1007/s11036-012-0368-0}
\BIBentrySTDinterwordspacing

\bibitem{she2019cross}
C.~She, Y.~Duan, G.~Zhao, T.~Q.~S. Quek, Y.~Li, and B.~Vucetic, ``{Cross-Layer
  Design for Mission-Critical IoT in Mobile Edge Computing Systems},''
  \emph{IEEE Internet of Things J., early access}, 2019.

\bibitem{ngo2013energy}
H.~Q. Ngo, E.~G. Larsson, and T.~L. Marzetta, ``{Energy and Spectral Efficiency
  of Very Large Multiuser MIMO Systems},'' \emph{IEEE Transactions on
  Communications}, vol.~61, no.~4, pp. 1436--1449, 2013.

\bibitem{luMIMOOverview2014}
L.~{Lu}, G.~Y. {Li}, A.~L. {Swindlehurst}, A.~{Ashikhmin}, and R.~{Zhang},
  ``{An Overview of Massive MIMO: Benefits and Challenges},'' \emph{IEEE
  Journal of Selected Topics in Signal Processing}, vol.~8, no.~5, pp.
  742--758, Oct 2014.

\bibitem{LISWaad2018}
M.~Jung, W.~Saad, Y.~Jang, G.~Kong, and S.~Choi, ``{Performance Analysis of
  Large Intelligence Surfaces (LISs): Asymptotic Data Rate and Channel
  Hardening Effects},'' \emph{arXiv preprint arXiv:1810.05667}, 2018.

\bibitem{zhang2015CRAN}
L.~{Liu} and R.~{Zhang}, ``{Optimized Uplink Transmission in Multi-Antenna
  C-RAN With Spatial Compression and Forward},'' \emph{IEEE Transactions on
  Signal Processing}, vol.~63, no.~19, pp. 5083--5095, Oct 2015.

\bibitem{Barbarossa2013}
S.~{Barbarossa}, S.~{Sardellitti}, and P.~{Di Lorenzo}, ``{Joint Allocation of
  Computation and Communication Resources in Multiuser Mobile Cloud
  Computing},'' in \emph{2013 IEEE 14th Workshop on Signal Processing Advances
  in Wireless Communications (SPAWC)}, June 2013, pp. 26--30.

\bibitem{Barbarossa2015}
S.~{Sardellitti}, G.~{Scutari}, and S.~{Barbarossa}, ``{Joint Optimization of
  Radio and Computational Resources for Multicell Mobile-Edge Computing},''
  \emph{IEEE Transactions on Signal and Information Processing over Networks},
  vol.~1, no.~2, pp. 89--103, June 2015.

\bibitem{jointKim2019}
J.~{Kim}, S.~{Park}, O.~{Simeone}, I.~{Lee}, and S.~{Shamai}, ``{Joint Design
  of Fronthauling and Hybrid Beamforming for Downlink C-RAN Systems},''
  \emph{IEEE Transactions on Communications}, pp. 1--1, 2019.

\bibitem{liu2019TwoScale}
A.~{Liu}, X.~{Chen}, W.~{Yu}, V.~K.~N. {Lau}, and M.~{Zhao}, ``{Two-Timescale
  Hybrid Compression and Forward for Massive MIMO Aided C-RAN},'' \emph{IEEE
  Transactions on Signal Processing}, vol.~67, no.~9, pp. 2484--2498, May 2019.

\bibitem{li2018minmax}
Q.~{Li}, J.~{Lei}, and J.~{Lin}, ``{Min-Max Latency Optimization for Multiuser
  Computation Offloading in Fog-Radio Access Networks},'' in \emph{2018 IEEE
  International Conference on Acoustics, Speech and Signal Processing
  (ICASSP)}, April 2018, pp. 3754--3758.

\bibitem{jointBckAcce2015}
O.~{Dhifallah}, H.~{Dahrouj}, T.~Y. {Al-Naffouri}, and M.~{Alouini}, ``{Joint
  Hybrid Backhaul and Access Links Design in Cloud-Radio Access Networks},'' in
  \emph{2015 IEEE 82nd Vehicular Technology Conference (VTC2015-Fall)}, Sep.
  2015, pp. 1--5.

\bibitem{de2019non}
E.~De~Carvalho, A.~Ali, A.~Amiri, M.~Angjelichinoski, and R.~W. Heath~Jr,
  ``{Non-Stationarities in Extra-Large Scale Massive MIMO},'' \emph{arXiv
  preprint arXiv:1903.03085}, 2019.

\bibitem{amiriXtra2018}
A.~{Amiri}, M.~{Angjelichinoski}, E.~{de Carvalho}, and R.~W. {Heath},
  ``{Extremely Large Aperture Massive MIMO: Low Complexity Receiver
  Architectures},'' in \emph{2018 IEEE Globecom Workshops (GC Wkshps)}, Dec
  2018, pp. 1--6.

\bibitem{BeyondMIMO2018}
S.~{Hu}, F.~{Rusek}, and O.~{Edfors}, ``{Beyond Massive MIMO: The Potential of
  Positioning With Large Intelligent Surfaces},'' \emph{IEEE Transactions on
  Signal Processing}, vol.~66, no.~7, pp. 1761--1774, April 2018.

\bibitem{Mar2014}
A.~O. {Martinez}, E.~{De Carvalho}, and J.~Ã. {Nielsen}, ``{Towards Very Large
  Aperture Massive MIMO: A Measurement Based Study},'' in \emph{2014 IEEE
  Globecom Workshops (GC Wkshps)}, Dec 2014, pp. 281--286.

\bibitem{CRanSurvey2014}
A.~{Checko}, H.~L. {Christiansen}, Y.~{Yan}, L.~{Scolari}, G.~{Kardaras}, M.~S.
  {Berger}, and L.~{Dittmann}, ``{Cloud RAN for Mobile Networks: A Technology
  Overview},'' \emph{IEEE Communications Surveys Tutorials}, vol.~17, no.~1,
  pp. 405--426, Firstquarter 2015.

\bibitem{yu2016alternating}
X.~Yu, J.~Shen, J.~Zhang, and K.~B. Letaief, ``{Alternating Minimization
  Algorithms for Hybrid Precoding in Millimeter Wave MIMO Systems},''
  \emph{IEEE Journal of Selected Topics in Signal Processing}, vol.~10, no.~3,
  pp. 485--500, April 2016.

\bibitem{sohrabi2016hybrid}
F.~Sohrabi and W.~Yu, ``{Hybrid Digital and Analog Beamforming Design for
  Large-Scale Antenna Arrays},'' \emph{IEEE Journal of Selected Topics in
  Signal Processing}, vol.~10, no.~3, pp. 501--513, April 2016.

\bibitem{addOffload2018}
W.~{Chen}, D.~{Wang}, and K.~{Li}, ``{Multi-user Multi-task Computation
  Offloading in Green Mobile Edge Cloud Computing},'' \emph{IEEE Transactions
  on Services Computing}, pp. 1--1, 2018.

\bibitem{offStrategy2019}
J.~{Zheng}, Y.~{Cai}, Y.~{Wu}, and X.~{Shen}, ``{Dynamic Computation Offloading
  for Mobile Cloud Computing: A Stochastic Game-Theoretic Approach},''
  \emph{IEEE Transactions on Mobile Computing}, vol.~18, no.~4, pp. 771--786,
  April 2019.

\bibitem{dong2019deep}
R.~Dong, C.~She, W.~Hardjawana, Y.~Li, and B.~Vucetic, ``{Deep Learning for
  Hybrid 5G Services in Mobile Edge Computing Systems: Learn from a Digital
  Twin},'' \emph{IEEE Transactions on Wireless Commun., early access}, 2019.

\bibitem{pcDeep2018}
W.~{Lee}, M.~{Kim}, and D.~{Cho}, ``{Deep Power Control: Transmit Power Control
  Scheme Based on Convolutional Neural Network},'' \emph{IEEE Communications
  Letters}, vol.~22, no.~6, pp. 1276--1279, June 2018.

\bibitem{decode2018}
F.~{Liang}, C.~{Shen}, and F.~{Wu}, ``{An Iterative BP-CNN Architecture for
  Channel Decoding},'' \emph{IEEE Journal of Selected Topics in Signal
  Processing}, vol.~12, no.~1, pp. 144--159, Feb 2018.

\bibitem{DeepUSVXu2019}
J.~{Xu}, P.~{Zhu}, J.~{Li}, and X.~{You}, ``{Deep Learning Based Pilot Design
  for Multi-user Distributed Massive MIMO Systems},'' \emph{IEEE Wireless
  Communications Letters}, pp. 1--1, 2019.

\bibitem{DeepSVKim2018}
K.~{Kim}, J.~{Lee}, and J.~{Choi}, ``{Deep Learning Based Pilot Allocation
  Scheme (DL-PAS) for 5G Massive MIMO System},'' \emph{IEEE Communications
  Letters}, vol.~22, no.~4, pp. 828--831, April 2018.

\bibitem{SunDeep2018}
H.~{Sun}, X.~{Chen}, Q.~{Shi}, M.~{Hong}, X.~{Fu}, and N.~D. {Sidiropoulos},
  ``{Learning to Optimize: Training Deep Neural Networks for Interference
  Management},'' \emph{IEEE Transactions on Signal Processing}, vol.~66,
  no.~20, pp. 5438--5453, Oct 2018.

\bibitem{luong2019applications}
N.~C. Luong, D.~T. Hoang, S.~Gong, D.~Niyato, P.~Wang, Y.-C. Liang, and D.~I.
  Kim, ``{Applications of deep reinforcement learning in communications and
  networking: A survey},'' \emph{IEEE Commun. Surveys Tuts., early access},
  2019.

\bibitem{LiuJointPower2015}
L.~{Liu}, S.~{Bi}, and R.~{Zhang}, ``{Joint Power Control and Fronthaul Rate
  Allocation for Throughput Maximization in OFDMA-Based Cloud Radio Access
  Network},'' \emph{IEEE Transactions on Communications}, vol.~63, no.~11, pp.
  4097--4110, Nov 2015.

\bibitem{learningSurvey2019}
C.~{Zhang}, P.~{Patras}, and H.~{Haddadi}, ``{Deep Learning in Mobile and
  Wireless Networking: A Survey},'' \emph{IEEE Communications Surveys
  Tutorials}, pp. 1--1, 2019.

\bibitem{kingma2014adam}
D.~P. Kingma and J.~Ba, ``{Adam: A Method for Stochastic Optimization},''
  \emph{arXiv preprint arXiv:1412.6980}, 2014.

\bibitem{SVDlow2018}
D.~{Zhang}, P.~{Pan}, R.~{You}, and H.~{Wang}, ``{SVD-Based Low-Complexity
  Hybrid Precoding for Millimeter-Wave MIMO Systems},'' \emph{IEEE
  Communications Letters}, vol.~22, no.~10, pp. 2176--2179, Oct 2018.

\bibitem{wang2008new}
Y.~Wang, J.~Yang, W.~Yin, and Y.~Zhang, ``{A New Alternating Minimization
  Algorithm for Total Variation Image Reconstruction},'' \emph{SIAM Journal on
  Imaging Sciences}, vol.~1, no.~3, pp. 248--272, 2008.

\bibitem{jain2017non}
P.~Jain, P.~Kar \emph{et~al.}, ``{Non-Convex Optimization for Machine
  Learning},'' \emph{Foundations and Trends{\textregistered} in Machine
  Learning}, vol.~10, no. 3-4, pp. 142--336, 2017.

\bibitem{bezdek2002some}
J.~C. Bezdek and R.~J. Hathaway, ``{Some notes on alternating optimization},''
  in \emph{AFSS International Conference on Fuzzy Systems}.\hskip 1em plus
  0.5em minus 0.4em\relax Springer, 2002, pp. 288--300.

\bibitem{bezdek2003convergence}
J.~Bezdek and R.~J. Hathaway, ``{Convergence of Alternating Optimization},''
  \emph{Neural, Parallel \& Scientific Computations}, vol.~11, no.~4, pp.
  351--368, 2003.

\bibitem{jain2013low}
P.~Jain, P.~Netrapalli, and S.~Sanghavi, ``{Low-Rank Matrix Completion using
  Alternating Minimization},'' \emph{Proceedings of the Forty-Fifth Annual ACM
  Symposium on Theory of Computing}, pp. 665--674, 2013.

\bibitem{marks1978general}
B.~R. Marks and G.~P. Wright, ``{A General Inner Approximation Algorithm for
  Non-convex Mathematical Programs},'' \emph{Operations research}, vol.~26,
  no.~4, pp. 681--683, 1978.

\bibitem{zappone2016EE}
A.~{Zappone}, L.~{Sanguinetti}, G.~{Bacci}, E.~{Jorswieck}, and M.~{Debbah},
  ``{Energy-Efficient Power Control: A Look at 5G Wireless Technologies},''
  \emph{IEEE Transactions on Signal Processing}, vol.~64, no.~7, pp.
  1668--1683, April 2016.

\bibitem{boyd2004convex}
S.~Boyd and L.~Vandenberghe, \emph{{Convex Optimization}}.\hskip 1em plus 0.5em
  minus 0.4em\relax Cambridge university press, 2004.

\bibitem{chollet2015keras}
F.~Chollet \emph{et~al.}, ``{Keras},'' \url{https://keras.io}, 2015.

\bibitem{tse2005fundamentals}
D.~Tse and P.~Viswanath, \emph{{Fundamentals of Wireless Communication}}.\hskip
  1em plus 0.5em minus 0.4em\relax Cambridge university press, 2005.

\bibitem{Yates1995PC}
R.~D. {Yates}, ``{A Framework for Uplink Power Control in Cellular Radio
  Systems},'' \emph{IEEE Journal on Selected Areas in Communications}, vol.~13,
  no.~7, pp. 1341--1347, Sep. 1995.

\end{thebibliography}

\end{document}